\documentclass[twocolumn]{aastex631}
\usepackage{amsmath}
\usepackage{multirow}
\usepackage{amssymb}
\usepackage{graphicx}
\usepackage{color}
\usepackage{float}
\usepackage{comment}

\newcommand{\remark}[1]{{\color{red}#1}} 

\definecolor{darkgreen}{rgb}{0,0.35,0}
\newcommand{\unit}[1]{\,\mathrm{#1}}  
\usepackage{CJK}

\begin{document}
\begin{CJK*}{UTF8}{gbsn}

\title{Accretion of AGN Stars under Influence of Disk Geometry II: The Adiabatic Regime and Runaway Collapse Induced by Self-gravity}

\shorttitle{3D Adiabatic Simulations of AGN-Embedded Stars}
\shortauthors{Chen et al.}

\author[0000-0003-3792-2888]{Yi-Xian Chen (陈逸贤)}
\affiliation{Department of Astrophysical Sciences, Princeton University,  4 Ivy Lane, Princeton, NJ 08544, USA}

\author{Yan-Fei Jiang (姜燕飞)}
\affiliation{Center for Computational Astrophysics, Flatiron Institute, New York, NY 10010, USA}

\author{Jeremy Goodman}
\affiliation{Department of Astrophysical Sciences, Princeton University,  4 Ivy Lane, Princeton, NJ 08544, USA}

\begin{abstract}
Accretion onto massive stars embedded in Active Galactic Nuclei (AGN) disks around supermassive black holes (SMBHs) is regulated to the stellar Eddington rate in the fast-diffusion or radiatively-efficient limit, 
$c/\tau > c_s$, where $\tau$ is the optical depth of the accretion flow and $c_s$ the sound speed. 
However, when the ambient density is sufficiently high, the opposite slow-diffusion limit applies. 
In this regime, accretion proceeds quasi-adiabatically and forms a hydrostatic circumstellar envelope (CSE) that stalls further mass inflow in the absence of self-gravity.
We perform 3D hydrodynamic simulations in the adiabatic limit to investigate the structure and evolution of such envelopes. 
For low thermal mass ratios, $q_{\rm th} \equiv M_\star/M_{\rm th}$ where $M_{\rm th}=c_s^3/(G\Omega)$ is the thermal mass, 
the CSE boundary smoothly matches the ambient disk entropy and density without forming a shock. 
 In contrast, when $q_{\rm th} \gg 1$, 
a strong shock develops at the envelope boundary, substantially increasing the entropy of the envelope and thereby regulating its structure and mass, $M_{\rm env}$.
In marginally self-gravitating disks with Toomre parameter $Q \sim 1$, 
we find that at sufficiently large $q_{\rm th}$ the envelope mass satisfies $M_{\rm env}/M_\star \gtrsim 1$. 
This condition is equivalent to stating that the post-shock material entering the envelope possesses lower radiation entropy than the characteristic stellar value, 
which triggers dynamical runaway growth on a dynamical timescale once envelope self-gravity is included in our simulations.
In realistic AGN disk environments with SMBH mass $\sim 10^8M_\odot$, runaway may occur close to the minimum self-gravitating radii and produce supermassive stars of 
$\sim 10^5M_\odot$. 
\end{abstract}

\section{Introduction}
\label{sec:intro}

Active Galactic Nuclei (AGN) host accretion disks around supermassive black holes (SMBHs; \citealt{Lyndenbell1969}).
Through gas drag and dissipative interactions, stars from the surrounding nuclear star cluster (NSC) on initially inclined, eccentric orbits can be captured into the disk and circularized  \citep{Artymowicz1993,MacLeod2020,WangYH2024}.
At large radii, 
AGN disks themselves become gravitationally unstable, leading to in-situ star formation through disk fragmentation \citep{Goodman2003,Levin2003,Jiang11,Chen2023}.
Once embedded in the dense, gas-rich midplane, AGN stars originated from both channels can undergo intense accretion, reaching masses and luminosities well above those of stars evolving in isolation.
The nucleosynthetic yields of stars embedded in AGN disks provide a natural explanation for the super-solar metallicities inferred from emission-line diagnostics 
in high-redshift quasars \citep{Hamann1999,Collin+Zahn1999,Nagao2006,Xu2018,Wang+2022,Lai+2022,Huang2023,Floris2024,Fryer2025}.
In parallel, 
the growth and subsequent dynamical evolution of massive stars in AGN disks can generate compact remnants that efficiently merge within the disk environment, 
offering a viable formation channel for the high-mass binary black hole mergers observed by the LIGO–Virgo Collaboration \citep{McKernan2012,McKernan2014,Tagawa2020a,Li2021,Samsing2022,Li2022,Chen2022,Epstein-Martin2024}.
Beyond these long-term outcomes, 
stellar populations in AGN disks may also give rise to distinctive electromagnetic transients, including proposed star-disk collision models for quasi-periodic eruptions \citep{Linial2023,Tagawa2023,JiangPan2025} and unusually energetic tidal disruption events \citep{Graham2025}.

1D studies using stellar evolution code MESA \citep{Cantiello2021,Dittmann2021,AliDib2023} have modeled the long-term evolution of stars embedded in AGN disks, 
adopting simplified prescriptions for gas accretion and mass loss and assuming spherical symmetry. 
In these models, 
stellar accretion is typically parameterized by a Bondi rate evaluated using characteristic AGN disk density and temperature, with radiative feedback 
from the stellar luminosity incorporated as an effective reduction of gravity by a factor scaling with $\lambda_\star = L_\star/L_{\rm Edd}$. 
Nevertheless, 
this treatment is insufficient to prevent a ubiquitous outcome of thermal runaway 
\citep{Fabj2025} at moderate ambient densities $\rho \gtrsim 10^{-11}$g/cm$^{-3}$ typical of AGN disks, 
namely that mass growth outpaces Kelvin-Helmholtz contraction and stars quickly gain masses of $\gtrsim 10^3-10^4 M_\odot$. 
The presence of 
a substantial population of such runaway stars would potentially produce frequent pair-instability supernovae and dramatically alter disk structure and emission.

In \citet{Chen2024}, 
we carried out radiation hydrodynamic simulations of accreting stellar envelopes in an isotropic background environment, 
demonstrating that radiative feedback operates in a more complex manner 
than assumed in existing 1D prescriptions. 
In the fast-diffusion regime 
where radiation forces efficiently counteract gravity, 
\textit{additional} feedback arising from radiation entropy and gravitational energy release within the accretion flow can suppress accretion by another one to two orders of magnitude, 
producing behavior analogous to Eddington-limited accretion onto black holes. 
This conclusion is further supported by vertically stratified simulations of circum-stellar accretion
that include tidal interactions and background shear velocity \citep{Chen2025}, 
and has been incorporated into a new suite of long-term MESA stellar evolution calculations \citep{Xu2025}, 
which demonstrate that such enhanced feedback effectively quenches thermal runaway. 

By contrast, 
in the extremely optically thick regime where the diffusion timescale exceeds the sound-crossing time ($c/\tau < c_s$, 
with $\tau$ denoting the optical depth of the accretion flow), 
radiation becomes tightly coupled to the gas in the form of an effective equation of state, 
and the accretion flow approaches an adiabatic limit.
In this regime, 
initial accretion without radiative feedback would rapidly inflate a thick envelope above the stellar surface within only a few dynamical timescales. 
However, this does not imply that the accumulated mass could be efficiently incorporated into the stellar core on dynamical timescales. 
Instead, 
the flow establishes an approximately isentropic envelope 
whose base pressure is insufficient to overcome the ram pressure barrier required for penetrating the stellar core, 
forcing further growth 
to proceed on thermal timescales, 
never really outpacing Kelvin-Helmholtz contraction. 
At the same time, 
pressure support in the hydrostatic envelope suppresses additional infall, 
rather than allowing accretion to proceed continuously in a rotating disk as in the fast-diffusion regime. 
Nevertheless, 
as noted by \citet{Chen2024}, 
if the stellar mass is high enough, 
the envelope itself may become self-gravitating, 
triggering 
a renewed runaway scenario via dynamical collapse of the envelope. 

In this paper, 
we re-investigate the slow-diffusion limit using three-dimensional, 
vertically stratified simulations in a geometry similar to that of \citet{Chen2025}. 
We show that, 
in realistic stratified AGN disk environments with background shear, 
reaching the self-gravitating envelope regime is generally difficult and requires not only high stellar mass but also a high enough ambient density, 
such that the background disk
must already lie in the marginally unstable regime with Toomre $Q\lesssim 1$.
The requirement is more stringent than the isotropic case considered in \citet{Chen2024} 
due to two key effects: 
i) the stellar sphere of influence is tidally truncated, 
effectively limiting envelope volume to be within the Hill radius even if the Bondi radius is much larger, 
and 
ii) gas entering the Hill sphere at supersonic shear velocities generates a strong standing shock that raises the entropy of the envelope, which raises the envelope density but more importantly
enhances pressure support against self-gravity. 
This paper is organized as follows:
In \S \ref{sec:setup}, we describe the numerical setup of our simulations.
In \S \ref{sec:results}, we present results from our primary suite of non-self-gravitating simulations and examine the relationship between the steady-state envelope mass and the stellar core mass. We also report results from a limited set of self-gravitating simulations and identify a restricted region of parameter space in which runaway accretion may occur. 
Finally, in \S \ref{sec:discussions}, 
we discuss the implications of the various outcomes in the slow-diffusion regime for the long-term evolution of AGN stars in high-density environments.

\section{3D Numerical Setup}
\label{sec:setup}

We use \texttt{Athena++} \citep{Stone2020} to perform hydrodynamic simulations with and without self-gravity. 
In a spherical polar coordinate system $(R, \theta, \phi)$ centered on the AGN star with mass $M_\star$, 
we solve hydrodynamic equations in the adiabatic limit with an Eddington equation of state (EOS) where the pressure is: 

\begin{equation}
    P = \rho c_s^2 = \dfrac{\rho \mathcal{R} T}{\mu} + \dfrac{aT^4}3,
    \label{eqn:eddington_quartic}
\end{equation}

and the energy is

\begin{equation}
    U =  \dfrac{3\rho \mathcal{R} T}{2\mu} + {aT^4}.
\end{equation}

Here $\mathcal{R}$ is the gas constant, 
and $\mu =0.6$ is the molecular weight. 
This EOS assumes that the gas and radiation are sufficiently well coupled to have a common temperature $T$,  and is implemented through the 
general EOS module of \texttt{Athena++} \citep{Coleman2020}. . 
The potential from both the central star $M_\star$ and the SMBH with $M_\bullet = 10^8 M_\odot$ located at 
$(R,\theta,\phi) = (r = 0.003\unit{pc},\, 0,\, \pi/2)$, 
as well as the indirect potential and Coriolis force of the frame rotating at 
$\Omega = \sqrt{GM_\bullet/r^3}$, 
are implemented following the procedure described in \citet{Chen2025}. We specifically use $r$ to denote distance from the SMBH in more global contexts, 
differentiating from $R$ as distance to the embedded star in the context of our local simulations. 
The simulation domain resolves the angles $\phi \in [0,2\pi]$ and $\theta\in [0,\pi/2]$ on uniform meshes with $N_\phi = 256$, $N_\theta = 64$, 
and a logarithmic radial grid ranging over
$R\in [50 R_\odot, R_{\rm out}]$ with $N_R=128$, 
where $R_{\rm out}$ takes the fiducial value of 2000$R_\odot$ which is considerably larger than the Hill radius $R_H$ in most of our simulations, where $q = M_\star/M_\bullet$ is the mass ratio. 
In the particularly self-gravitating simulation presented in \S \ref{sec:sg} with runaway growth of the envelope mass as well as $R_H$ of the star-envelope system, 
we perform a convergence test with a larger $R_{\rm out} = 3000 R_\odot$. 

Similar to \citet{Chen2025}, 
the $\phi$ boundaries are periodic and $\theta$ boundaries are reflective. 
The radial outer boundary condition is set by an AGN disk model with midplane density $\rho_0$ and temperature $T_0$, 
with a vertical distribution 
following a $P\propto \rho^{4/3}$ polytrope with constant adiabatic radiation-to-gas-pressure ratio $\Pi_0 = a_{\rm rad} T_0^3\mathcal{R}/3\mu \rho_0$. 
The associated sound speed $c_{s,0}$ defined by Equation \eqref{eqn:eddington_quartic} also establishes a characteristic scale height $H = c_{s,0}/\Omega$.
The fixed velocity at $R_{\rm out}$ is set by the background Keplerian shear in the corotating frame, 
where all velocity vectors point normal to the direction of the SMBH host. 
This treatment gradually injects the disk gas into the simulation domain once the simulation starts evolving. 

For the inner boundary condition in $R$, 
the radiation hydrodynamic simulations in \citet{Chen2025} use a fixed $\rho, T$ profile and set the velocity to zero. 
Those simulations in the radiatively efficient regime
start with a realistic stellar profile extracted from \texttt{MESA} near the radial inner boundary 
and evolve the system to a state in which radiative diffusion provides sustained feedback, 
while low-$\dot{M}$ accretion feeds 
a stellar envelope that gains negligible mass and remains close to its initial configuration.
However, 
in this paper we focus on radiatively-inefficient regimes characterized by significantly higher $\rho_0$  
and expect a fully coupled state for radiation and gas due to high optical depth. 
Specifically,
we consider the highest $\rho_0$ values limited by the onset of self-gravity regulation for order-unity Toomre parameter of the disk:

\begin{equation}
    Q \equiv  \dfrac{\Omega^2}{2\pi G\rho_0}\gtrsim 1
\end{equation}

which yields $\rho \sim 10^{-8}$g/cm$^{-3}$ for our $\Omega$ value. 
As we will emphasize below, 
the strong opacity and long diffusion timescales in this setup are consistent with the slow diffusion regime, 
which justifies and necessitates the use of adiabatic simulations that do not require full radiative transfer.

As suggested in \citet{Chen2024} and in studies of circumplanetary flows \citep[e.g.][]{Fung2019},
in this regime the flow rapidly establishes an approximately isentropic envelope near the star, 
whose hydrostatic structure is set primarily by the background disk conditions $(\rho_0, T_0)$ rather than by the intrinsic stellar properties.
As a result, any initially imposed stellar envelope is quickly erased by mass accumulation as this quasi-steady configuration forms.
Motivated by this behavior, and to enable a systematic exploration over a wide range of $M_\star$, we model the star as a point-mass potential, 
without prescribing an explicit initial stellar structure: i.e., the original mass of the star is assumed to lie interior to the innermost radial grid point of the computational domain ($R_\mathrm{in}=50\,R_\odot$).
This choice greatly simplifies long-term parameter surveys by removing the need to construct a dedicated \texttt{MESA} stellar model for each stellar mass.
For simplicity, we adopt a reflecting inner radial boundary condition, 
implicitly assuming that the external envelope remains largely decoupled from the deep stellar interior.
This approximation is valid provided that the ram pressure of the accreting material remains 
insufficient to penetrate the stellar core on dynamical timescales, and that the stellar thermal adjustment timescale is much longer than the dynamical evolution of the envelope, 
so that responses deep within the inner core do not significantly influence the envelope structure.

By varying $M_\star$ and the density and temperature of the disk midplane, 
we will be able to follow convention 
of the circum-planetary disk literature and define a dimensionless parameter

\begin{equation}
    q_{\rm th} = \dfrac{R_B}{H} = \dfrac{M_\star}{M_\mathrm{th}}
\end{equation}

where the stellar Bondi radius is defined as 
\begin{equation}
    R_B = GM_\star/c_{s,0}^2,
\end{equation}

and the thermal mass of the disk is 

\begin{equation}
    M_\mathrm{th}=c_{s,0}^3/G\Omega.
\end{equation}

When $q_{\rm th} \lesssim 1$ and $H \gtrsim R_H \gtrsim R_B$, 
the predominantly vertical shear flow enters an approximately spherical envelope of characteristic size $R_B$ with sub-or trans-sonic velocities without a shock \citep[e.g.][]{Fung2019,Paardekooper2023}.
By contrast, as we will show below, in the superthermal regime where $q_{\rm th} > 1$ and $R_B \gtrsim R_H \gtrsim H$, 
tidal forces from the SMBH and the accretion shock generated as the supersonic background shear flow penetrates the envelope, 
now with a characteristic radius $\sim R_H$, 
play a dominant role in shaping the envelope structure and thermodynamics.

The full suite of models explored in this work is summarized in Table \ref{tab:parameters}.
The primary control parameters are $\rho_0$ and $T_0$, which determine the background disk properties through the derived quantities $\Pi_0$, $Q$, and $H$. We also show the effective adiabatic index $\gamma_0$ that is a function of $\Pi_0$ only for the Eddington EOS \citep{Mihilas1984,GoodmanTan2004}.
The stellar mass $M_\star$ sets the strength of tidal interactions, from which the Hill radius $R_H$, Bondi radius $R_B$, and thermal mass ratio $q_{\rm th}$ are derived.
Unless otherwise stated, all simulations are evolved for $\sim 50$ orbital periods to reach a quasi-steady state.




\section{Results}
\label{sec:results}

\begin{table*}
\centering
\begin{tabular}{ccccccccccc|c}
\hline
Model 
& $\rho_{0}$ (g cm$^{-3}$) 
& $Q$ 
& $T_{0}$ (K) 
& $\Pi_{0}$ 
& $\gamma_0$
& $H(R_\odot)$ 
& $M_\star (M_\odot)$ 
& $R_B(R_\odot)$ 
& $R_H(R_\odot)$ 
& $q_{\rm th}$ 
& Figure \\
\hline
\texttt{D1e-8T8e4m10}
& $10^{-8}$ & 3.9 & $8\times 10^4$ & 0.94 & 1.43 & 518.0 & 10
& 890.9 & 430.3 & 1.72
& \ref{fig:T8e4_rho1e-8_qth1},\ref{fig:T8e4_rho1e-8_qth1_radial},\ref{fig:entropy_T80000},\ref{fig:time_evo_T80000},\ref{fig:summary_T80000},\ref{fig:lum_tcool_T80000} \\
\hline
\texttt{D1e-8T8e4m20}
& -- & -- & -- & -- & -- & -- & 20
& 1781.8 & 542.1 & 3.44
& \ref{fig:entropy_T80000},\ref{fig:time_evo_T80000},\ref{fig:summary_T80000},\ref{fig:lum_tcool_T80000} \\
\hline
\texttt{D1e-8T8e4m50}
& -- & -- & -- & -- & -- & -- & 50
& 4454.7 & 735.9 & 8.59
& \ref{fig:T8e4_rho1e-8_qth8},\ref{fig:T8e4_rho1e-8_qth1_radial},\ref{fig:entropy_T80000},\ref{fig:time_evo_T80000},\ref{fig:summary_T80000},\ref{fig:lum_tcool_T80000} \\
\hline
\texttt{D1e-8T8e4m100}
& -- & -- & -- & -- & --& -- & 100
& 8909.5 & 927.1 & 17.2
& \ref{fig:time_evo_T80000},\ref{fig:summary_T80000},\ref{fig:lum_tcool_T80000} \\
\hline
\texttt{D1e-8T4e4m1}
& -- & -- & $4\times 10^4$ & 0.12  & 1.56& 278.4 & 1
& 308.4 & 199.7 & 1.11
& \ref{fig:summary_T40000} \\ 
\hline
\texttt{D1e-8T4e4m5}
& -- & -- & -- & --& -- & -- & 5
& 1541.9 & 341.5 & 5.54
&  \ref{fig:summary_T40000}\\
\hline
\texttt{D1e-8T4e4m10}
& -- & -- & -- & --& -- & -- & 10
& 3083.8 & 430.3 & 11.1
&   \ref{fig:summary_T40000}\\
\hline
\texttt{D1e-8T4e4m50}
& -- & -- & -- & --& -- & -- & 50
& 15418.8 & 735.9 & 55.4
&   \ref{fig:summary_T40000}\\
\hline
\texttt{D1e-8T4e4m100}
& -- & -- & -- & --& -- & -- & 100
& 30837.6 & 926.0 & 110.7
&  \ref{fig:summary_T40000}\\
\hline
\texttt{D1e-8T4e4m100\_sg}
& -- & -- & -- & -- & -- & --
& -- & -- & -- & --
&   \ref{fig:time_evo_sg}\\
\hline
\texttt{D3e-8T4e4m100\_sg}
& $3\times 10^{-8}$ & 1.3 & -- & 0.039 & 1.61 & 123.4 & --
& 33143.8 & -- & 268.6
&   \ref{fig:time_evo_sg}\\
\hline
\end{tabular}
\caption{List of adiabatic model names and input parameters used in our simulations. 
The SMBH mass is fixed at $M_\bullet = 10^8 M_\odot$ with a distance of $r = $0.003\,pc and orbital period of $2\pi/\Omega = 1.63$\,yr. 
\texttt{sg} indicates runs with self gravity. }
\label{tab:parameters}
\end{table*}

\subsection{The Isentropic Envelope}
\label{sec:isentropic}
\begin{figure*}
    \centering
    \includegraphics[width=1.0\textwidth]{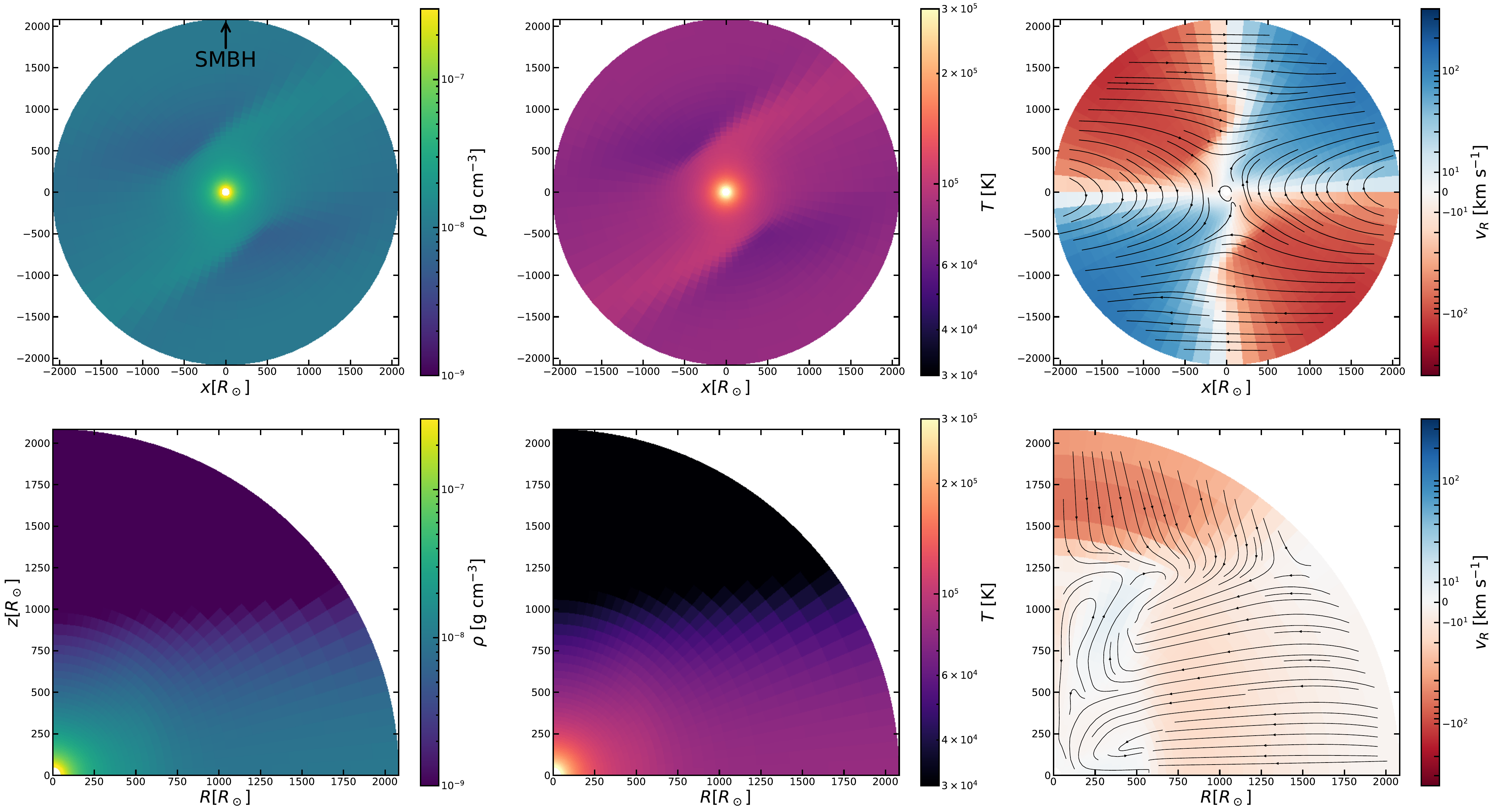}
    \caption{The marginally superthermal simulation \texttt{D1e-8T8e4m10} in quasi-steady state. The upper panel shows midplane distribution of density, temperature, infall velocity and 2D velocity streamlines (with $y$ axis along the direction of the SMBH host), 
    while the lower panels shows the corresponding quantities on the $R-\theta$ or $R-z$ plane averaged over azimuth $\phi$.}
\label{fig:T8e4_rho1e-8_qth1}
\end{figure*}

Figure~\ref{fig:T8e4_rho1e-8_qth1} presents the density, temperature, 
and radial velocity structure of run \texttt{D1e-8T8e4m10} with background $T = 8\times 10^4$K, 
shown in the midplane (upper panels) and in the azimuthally averaged vertical slices (lower panels). 
For this stellar mass, 
the system is marginally superthermal ($q_{\rm th}=1.72$), 
and a quasi-spherical envelope forms within $\sim 500R_\odot$, 
comparable to both the Hill radius and the local scale height.
The velocity amplitudes inside the envelope are substantially reduced relative to the surrounding flow, 
indicating a predominantly pressure-supported, 
nearly hydrostatic structure.

\begin{figure*}
    \centering
    \includegraphics[width=1.0\textwidth]{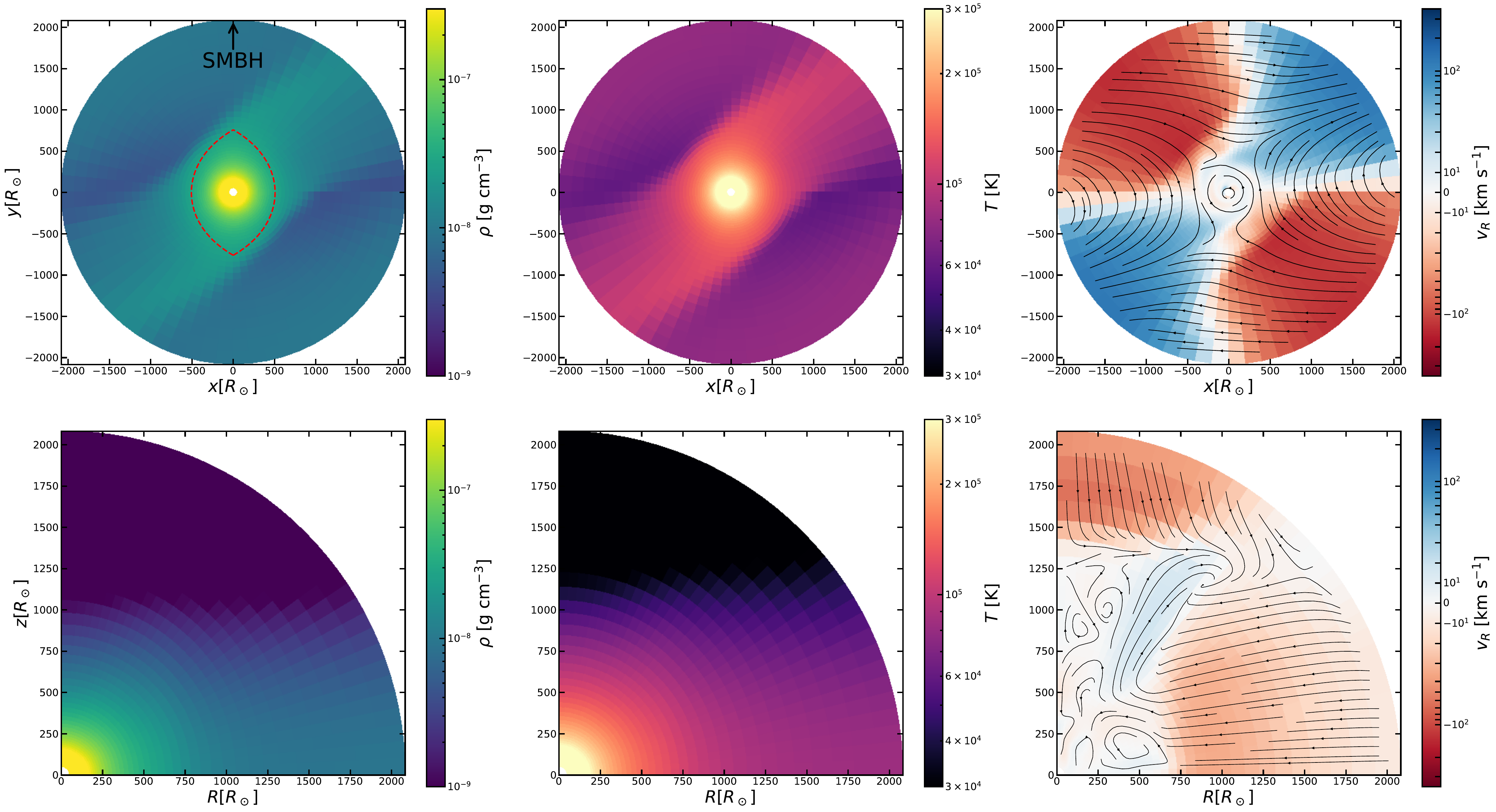}
    \caption{Similar to Figure \ref{fig:T8e4_rho1e-8_qth1} but for the superthermal simulation \texttt{D1e-8T8e4m50}. 
    The shock at the envelope boundary is much more conspicuous and is nearly tangential to the Hill sphere 
    (with the locus of the Roche lobe indicated by the dashed red line in upper left panel).
    }
\label{fig:T8e4_rho1e-8_qth8}
\end{figure*}

In turn, Figure~\ref{fig:T8e4_rho1e-8_qth8} shows the final snapshot of run \texttt{D1e-8T8e4m50}, 
which has a stellar mass five times larger than that in Figure~\ref{fig:T8e4_rho1e-8_qth1} and is plotted using the same color scales.
The envelope is correspondingly more extended, 
with substantially higher central density and temperature.
A pronounced shock is sustained at the envelope boundary, 
acting as an effective barrier that stalls the background shearing flow. 
Gas entering the envelope undergoes compression and heating across this shock, 
leading to significant enhancements 
in both density and temperature relative to the $q_{\rm th}\sim 1$ case 
\footnote{This shock is different from shocks produced by the steepening of density waves launched by sub-thermal companions ($q_{\rm th} \ll 1$) 
which occur only after the wave has propagated a distance of 
$ q_{\rm th}^{-2/5} H \gg H \gg R_H$ 
from the perturber \citep{Goodman2001}, 
as well as transient spiral shocks that will be discussed in \S \ref{sec:rotation}.}.

\begin{figure*}
    \centering
\includegraphics[width=1.0\textwidth]{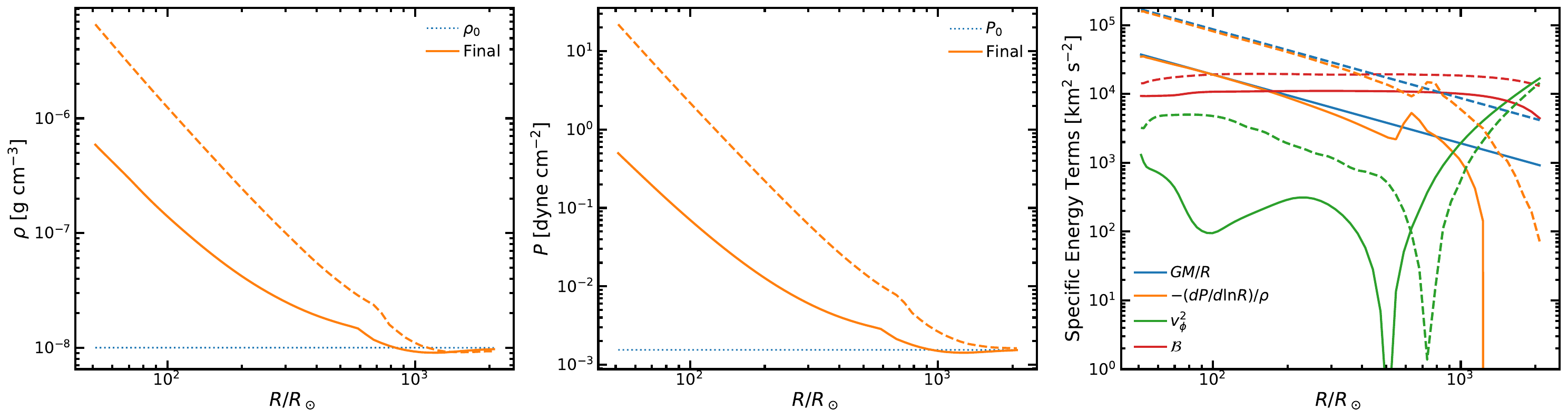}
    \caption{$\phi$-averaged midplane profiles for simulations \texttt{D1e-8T8e4m10} (solid lines) and \texttt{D1e-8T8e4m50} (dashed lines) in quasi-steady state. 
    The rightmost panel shows a variety of specific-energy terms [dimensions of (velocity)$^2$.]  }
\label{fig:T8e4_rho1e-8_qth1_radial}
\end{figure*}

Figure~\ref{fig:T8e4_rho1e-8_qth1_radial} shows the $\phi$-averaged midplane profiles of density and pressure (left and middle panels), 
together with several terms relevant to the radial force balance (right panel), 
for both simulations in their quasi-steady states. 
We explicitly demonstrate the dominant support against the stellar gravitational potential. 
In particular, the outward pressure force per unit mass,

\begin{equation}
    -\dfrac{1}{\rho } \dfrac{d P}{d R}
\end{equation}
nearly cancels the inward gravitational acceleration,

\begin{equation}
    - \dfrac{d\Phi_\star}{dR} = - \dfrac{GM_\star}{R^2}
\end{equation}
over a broad range of radii. 
By contrast, 
the azimuthal velocity remains strongly sub-Keplerian, 
and the centrifugal term $v_\phi^2/R$ contributes only a minor fraction of the required radial support.
Near the Hill radius, 
$R \sim R_H$, the azimuthally averaged $v_\phi$ drops to zero, 
marking a separatrix where coherent circumstellar rotation terminates and the flow transitions smoothly into the 
background Keplerian shear.

A more general way of examining the binding of the envelope is to plot the specific Bernoulli parameter, 
which can be approximated as 

\begin{equation}
    \mathcal{B} = \dfrac{v^2}{2} + \mathcal{H} + \Phi - \Phi_H,
\end{equation}
in which $\Phi$ is the summation of stellar, SMBH and indirect potentials \citep{Chen2025}, 
and $\Phi_H$ is a reference value defined at the L1 point $(R, \theta,\phi) = (R_H, 0, \pi/2)$. 
When $\mathcal{B}>0$, 
the energetic content of the gas may allow it to become unbound from the star, 
with the enthalpy defined as

\begin{equation}
    \mathcal{H} =  \dfrac{4}{3}\dfrac{a T^4}{\rho} +  \dfrac{5}{2}\dfrac{\mathcal{R}T}{\mu}, 
\end{equation}
for the Eddington EOS. 
We also plot the azimuthally averaged midplane radial profile for $\mathcal{B}$ in the right panel of Figure \ref{fig:T8e4_rho1e-8_qth1_radial}. 
Before azimuthal averaging,

\begin{equation}
    \Phi - \Phi_{H} \approx  -\dfrac{GM_\star}{R} - \dfrac32 \Omega^2 y^2 + \dfrac{GM_\star}{R_H}  + \dfrac32 \Omega^2 R_H^2\,,
\end{equation}
$y$ being measured from the star toward the SMBH (equivalent to radial coordinate $-x$ in a conventional shearing box). For $R < R_H$, the stellar potential dominates this expression.
In both simulations presented,
$\mathcal{B}$ remains positive over much of the envelope. 
This is probably because 
during envelope formation, 
most of the accreted gas originates from regions $y \sim R_H$ where the Bernoulli parameter is initially positive due to background Keplerian shear.\footnote{More explicitly, near the midplane the region $y\in [0.7, 3.1] R_H$ has $\mathcal{B}(t = 0)>0$ even when disk enthalpy is neglected (note $h\ll |\mathcal{B}|$ when $q_{\rm th} \gg 1$). 
Beyond $ \sim 3.1 R_H$ the term $- 3 \Omega^2 y^2/2$ dominates $\Phi$ and establishes a tidal barrier for accretion}. 
Because adiabatic shocks reduce the Bernoulli parameter only modestly, 
it is approximately conserved as the gas enters and accumulates within the envelope, leaving much of the material formally unbound unless additional cooling is introduced. 
Nevertheless, 
the condition $\mathcal{B} \ll -\Phi_\star$ provides a conservative criterion for effective binding, 
since material in this regime cannot easily overcome the stellar gravitational potential. 
We therefore still identify gas within $\sim R_H$, 
predominantly pressure-supported, 
as the circumstellar envelope 
discussed in the remainder of this work. 

\subsection{Entropy and mass of the envelope}
\label{sec:entropy_mass}
Even when the envelope is only marginally bound (and more formally bound at larger $q_{\rm th}$), 
it remains meaningful to quantify the mass contained within it.
One might attempt to estimate the envelope structure by integrating inward from the background boundary conditions $(\rho_0, T_0)$ under the assumption of constant entropy $\sim S_0(\rho_0, T_0)$.
In practice, 
however, this approach is complicated by the presence of a shock at the envelope boundary.
Across this shock, both the density and pressure exhibit discontinuous jumps, 
such that the characteristic entropy inside the envelope, $S'$, differs from the background midplane entropy $S_0$.
We therefore distinguish between these two entropies and define the radiation and gas entropy as (in units of $\mathcal{R}/\mu$)

\begin{equation}
    S_{\rm rad} = \frac{4 P_{\mathrm{rad}}}{P_{\mathrm{gas}}} =4\Pi,\quad S_{\rm gas} = \frac{1}{2} \ln \left(\frac{P_{\mathrm{rad}}}{P_{\mathrm{gas}} \tilde{\rho}}\right),
\end{equation}
where $\tilde{\rho} = \rho/10^{-7}$g cm$^{-3}$ with the normalization constant being our density code unit. 
Choosing a different normalization alters $S_\mathrm{gas}$ by an additive constant, and therefore does not affect entropy differences between different $(\rho, T)$ states.

\begin{figure*}
    \centering
    \includegraphics[width=1.0\textwidth]{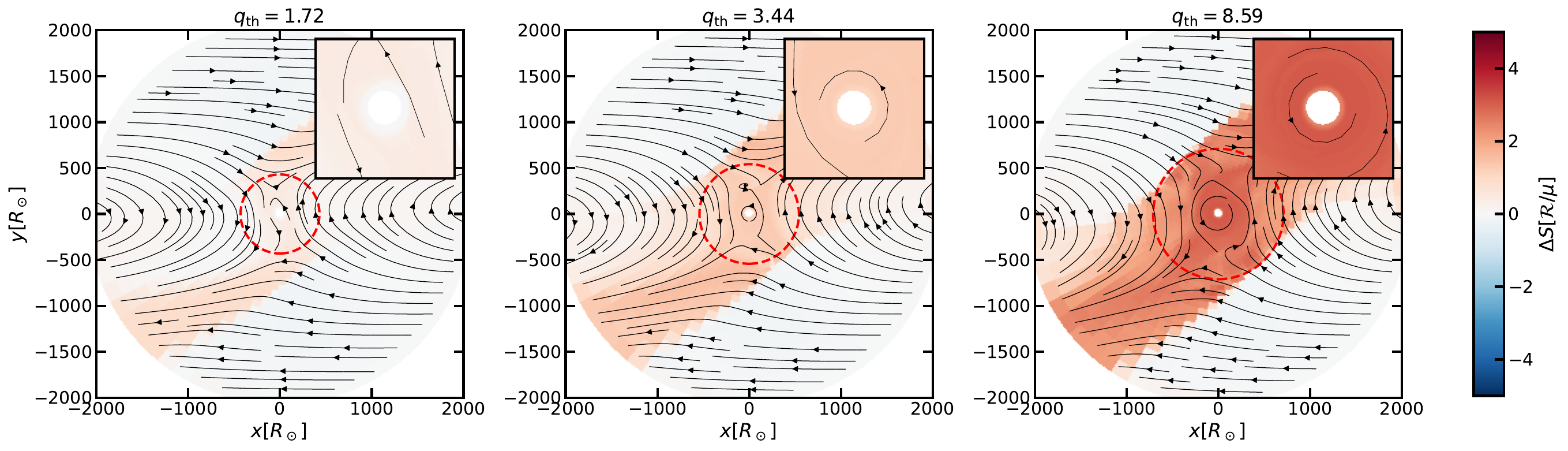}
    \includegraphics[width=1.0\textwidth]{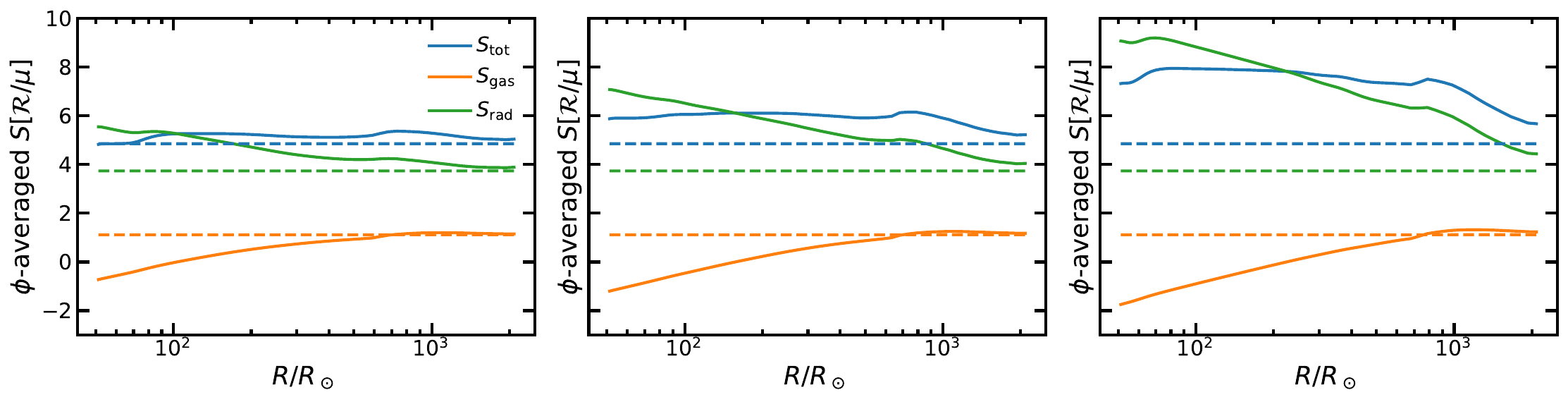}
    \caption{Upper panel: entropy distribution on the midplane for a set of simulations \texttt{D1e-8T8e4m10}, \texttt{D1e-8T8e4m20}, \texttt{D1e-8T8e4m50}with $T_0 = 80000$K and $\rho_0 = 10^{-8}$ g/cm$^{-3}$ in quasi-steady state. Dashed red circles indicate $R_H$.
    Lower panel: 
    $\phi$-averaged radial entropy profiles for different components in the midplane 
    (and the disk/initial values in dashed lines).}
\label{fig:entropy_T80000}
\end{figure*}

\begin{figure}
    \centering
    \includegraphics[width=0.43\textwidth]{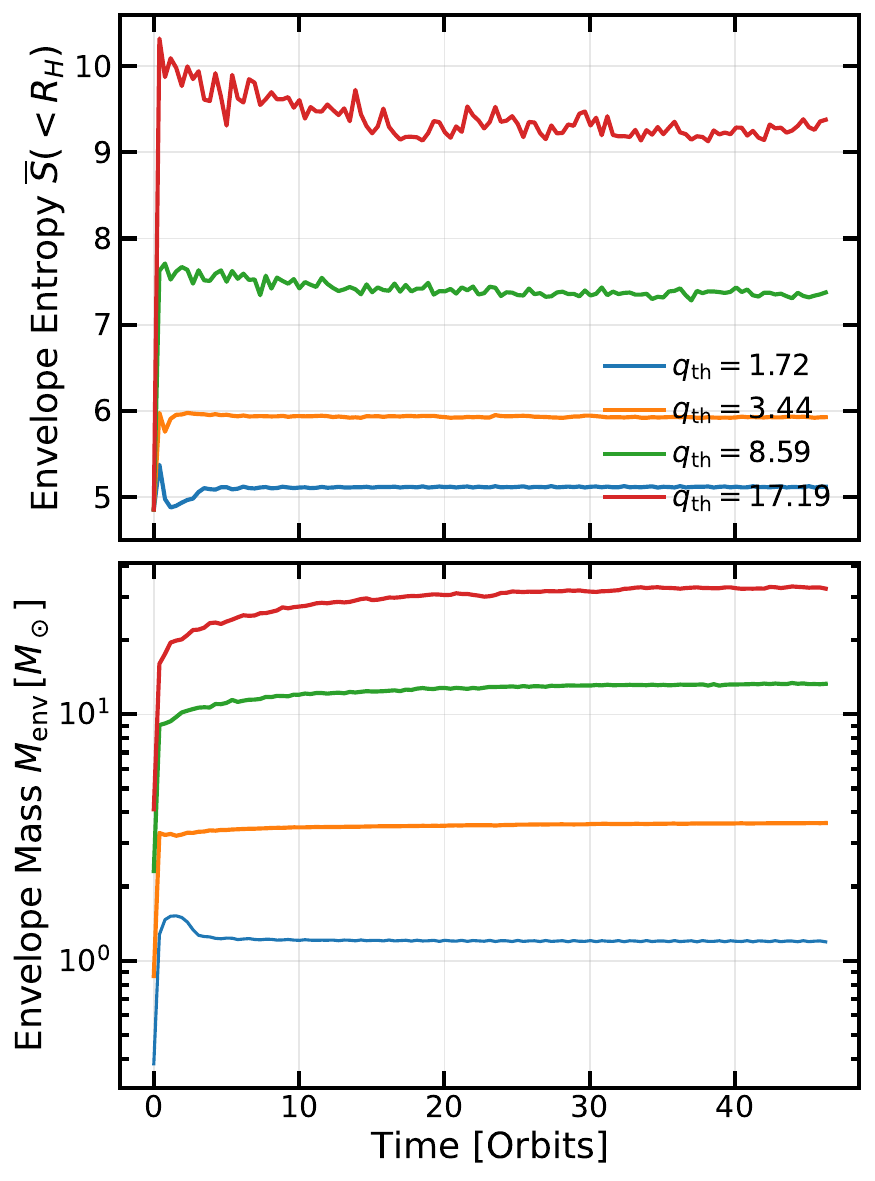}
    \caption{evolution of entropy and mass of envelope towards steady state for a set of simulations \texttt{D1e-8T8e4m10}, \texttt{D1e-8T8e4m20}, \texttt{D1e-8T8e4m50}, \texttt{D1e-8T8e4m100} with $T_0 = 80000$K and $\rho_0 = 10^{-8}$ g/cm$^{-3}$. }
\label{fig:time_evo_T80000}
\end{figure}

Figure~\ref{fig:entropy_T80000} shows the 2D midplane distribution of entropy (upper panels) and its $\phi$-averaged radial profiles (lower panels), for runs \texttt{D1e-8T8e4m10}, \texttt{D1e-8T8e4m20}, and \texttt{D1e-8T8e4m50} in quasi-steady state. 
All cases share the same background disk conditions, 
with the reference entropy $S_0=4.8$ computed from $\rho_0$ and $T_0$.
In the low-$q_{\rm th}$ regime, 
the entropy within the envelope exceeds $S_0$ only modestly, 
with the largest enhancement occurring near the envelope boundary at $R_H$ in the $y$ direction towards the SMBH. 
The material with moderate entropy excess is mostly transported away by shear flow, 
leaving the innermost regions of the circumstellar envelope, well separated from the shock, nearly isentropic at $S_0$.
As $q_{\rm th}$ increases, 
the shock strengthens and the associated entropy jump becomes substantial. 
In this regime, 
shock heating influences the entire envelope, 
producing an extended entropy plateau,\footnote{During early transient phases the envelope is hotter and exhibits a negative radial entropy gradient; 
it subsequently cools as convection carries energy outward.}
such that the envelope is effectively ``anchored" onto post-shock 
density $\rho'$, pressure $P'$ and entropy $S'$ rather than the disk background midplane values. 
For a measurement of entropy representative of the envelope, 
we calculate the mass-weighted average entropy within $R_H$ (red circles in upper panel of Figure \ref{fig:entropy_T80000}),

\begin{equation}
    \overline{S}(<R_H): = \dfrac{\int_{<R_H} \rho S dV}{\int_{<R_H} \rho  dV} = \dfrac{\int_{<R_H} \rho S dV}{M_{\rm env}},
\end{equation}
which is dominated by the post-shock characteristic entropy $S'$ rather than by the background disk value $S_0$.
The temporal evolution of both the mass-weighted entropy $\overline{S}(<R_H)$ and the enclosed envelope mass $M_{\rm env}$ is shown in Figure~\ref{fig:time_evo_T80000} for all simulations with $T_0 = 8\times10^4\,$K.
In all cases, including the most massive, strongly superthermal runs, 
both quantities converge to well-defined steady values after $\sim 30$ orbits, 
demonstrating that the envelopes reach quasi-steady states on the simulated timescales. 
The entropy excess $\overline{S}(<R_H)-S_0$ increases monotonically with $q_{\rm th}$, reflecting the growing importance of shock heating at the envelope boundary.

The presence of a shock should be
a natural outcome when the embedded companion is superthermal, 
as the background shear flow impinging on the Hill sphere becomes supersonic and thus satisfies the conditions for shock formation even in the adiabatic limit.
A similar effect may be expected in studies of circumplanetary envelopes (CPEs). However, 
most existing \textit{adiabatic} CPE simulations do not study this thermal mass regime in details. 
For example, \citet{Fung2019} presented 3D adiabatic simulations with $q_{\rm th}=1$ and an ideal-gas equation of state with $\gamma=7/5$, 
and reported no prominent shock features. As a result, 
shock formation has generally not been emphasized in the CPE literature.
We are aware of one exception: \citet{Bethune2019} presented two-dimensional adiabatic simulations with $q_{\rm th}=2$ and $\gamma=7/5$, 
which do exhibit an entropy jump consistent with shock heating, 
although the physical origin and implications of this feature were not quantified in detail (see their \S 5).

We now outline a simple framework for estimating the post-shock entropy $S'$, which serves as a proxy for the characteristic envelope entropy $\overline{S}(<R_H)$ that we can later use to calculate the envelope mass.
While this approach can be specialized to constant-$\gamma$ EOS as a direct supplement to existing CPE studies, 
we here formulate for a more general EOS.
As a first approximation, 
we assume that the pre-shock flow velocity is of order the Hill velocity $ v_H = \Omega R_H$, 
such that the Mach number satisfies 
$\mathcal{M} \sim v_H/c_{s,0}$. 
This allows the Mach number to be related to the thermal mass ratio as

\begin{equation}
    \mathcal{M}(q_{\rm th}) \approx (m q_{\rm th})^{1/3}
\end{equation}
where $m$ is an order-unity coefficient encapsulating geometric and flow-dependent factors. 


With this definition, 
the Rankine-Hugoniot conditions yield a closed set of equations for the compression ratio $r \equiv \rho'/\rho$ and post-shock pressure $P'$:

\begin{equation}
\left\{
\begin{aligned}
    P' - P_0 &= \mathcal{M}^2  P_0 (1 - 1/r) \\
    \mathcal{H}(r\rho_0, P') - \mathcal{H}(\rho_0, P_0) &= \frac{\mathcal{M}^2  P_0}{2 \rho_0} (1 - 1/r^2)
\end{aligned}
\right.
\label{eqn:compression_ratio}
\end{equation}
which can be solved iteratively to obtain solutions for both $P'/P_0$ and $r$. For ideal EOS we obtain the classic result

\begin{equation}
    r =\frac{(\gamma+1) \mathcal{M}^2}{(\gamma-1) \mathcal{M}^2+2}
    \label{eqn:compression_rho}
\end{equation}

\begin{equation}
    \frac{P'}{P_0}=\frac{2 \gamma \mathcal{M}^2-(\gamma-1)}{\gamma+1}
    \label{eqn:compression_P}
\end{equation}

\begin{figure}
    \centering
    \includegraphics[width=0.45\textwidth]{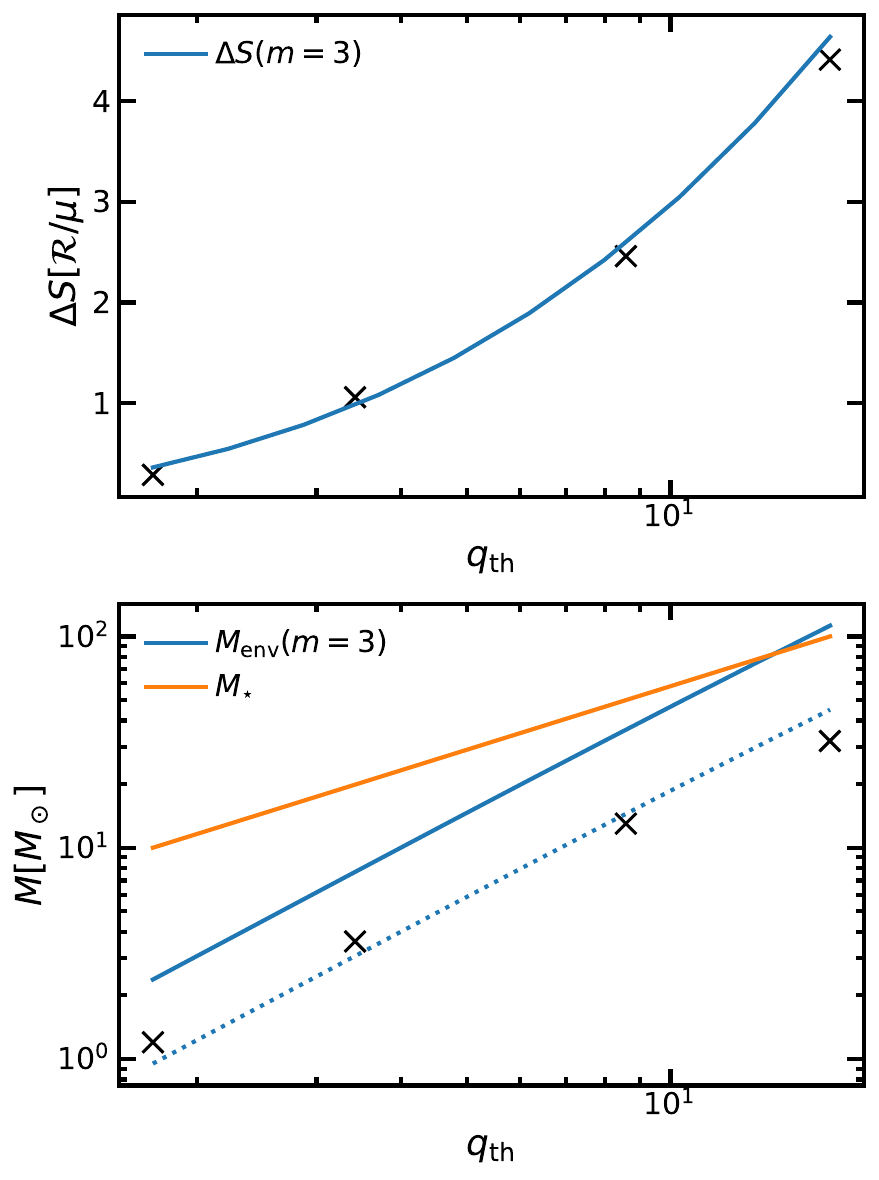}
    \caption{Simulation measurements (markers) 
    and physically-informed analytical fits (curves) for the entropy jump and circumstellar envelope mass, 
    evaluated for background disk $\rho_0 = 10^{-8}\mathrm{g/cm^{3}}$ and $T_0 = 8\times10^4\mathrm{K}$.}
\label{fig:summary_T80000}
\end{figure}

The upper panel of Figure~\ref{fig:summary_T80000} shows the theoretically predicted entropy jump, $\Delta S \equiv S' - S_0$, as a function of $q_{\rm th}$ for $\rho_0 = 10^{-8},\mathrm{g,cm^{-3}}$ and $T_0 = 8\times10^4,\mathrm{K}$, evaluated for $m=3$ using the generalized equation of state. 
The black markers indicate direct measurements from the simulations, providing an empirical calibration of the prefactor $m \approx 3$.

Following the procedure outlined above, we can also estimate the mass of a non-self-gravitating envelope as a function of the pre-shock Mach number $\mathcal{M}$, 
or equivalently the stellar mass $M_\star$ (or thermal mass ratio $q_{\rm th}$).
The lower panel of Figure \ref{fig:summary_T80000} shows the predicted mass of an isentropic envelope with boundary conditions $\rho(R_H)=\rho'$ and $S \equiv S'$, 
evaluated for $m=3$ (solid curve).
The direct analytical estimate systematically overpredicts the envelope mass relative to the simulations, 
likely because the idealized model assumes a spherical envelope, 
whereas the simulated envelopes are vertically stratified and increasingly rarefied at high altitudes. 
Accounting for this geometric effect, an empirical normalization of $\sim 40\%$ of the analytical value (dotted curve) provides a much closer match to the simulation measurements (black crosses).
The envelope mass generally scales super-linearly with stellar mass. 
Consequently, at sufficiently large $M_\star$, the envelope mass becomes comparable to the stellar mass itself, 
indicating that self-gravity must become dynamically important for the most massive embedded stars.
However, in the asymptotic limit of large $q_{\rm th}$, the scaling approaches $M_{\rm env} \sim M_\star / Q$, 
up to at most an additional logarithmic correction, which implies that runaway is not inevitable at arbitrarily large $q_{\rm th}$: if the disk Toomre parameter $Q$ is sufficiently large, 
the envelope remains pressure-supported and self-gravity cannot drive unlimited growth, at least not without cooling. 
We elaborate on this limiting behavior in \S \ref{sec:general_runaway}.

\subsection{Diffusion and Dynamical Timescales}

\begin{figure}
    \centering
    \includegraphics[width=0.45\textwidth]{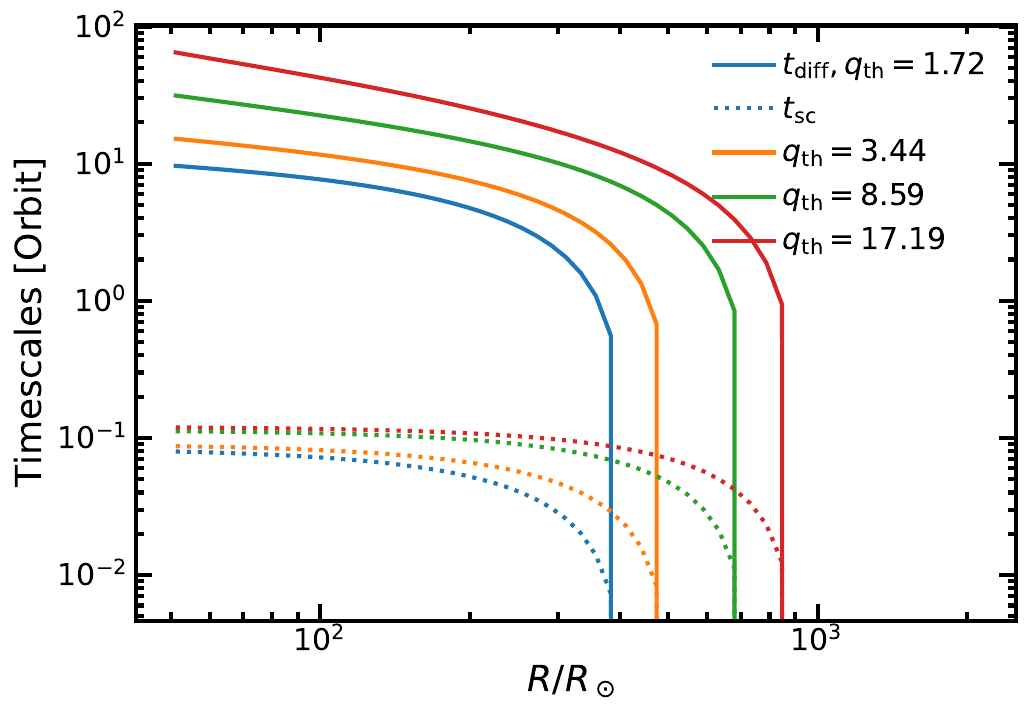}
    \caption{Radial profiles of the thermal diffusion timescale $t_{\rm diff}$ (solid lines) and sound-crossing time $t_{\rm sc}$ (dotted lines) for simulations with different thermal mass ratios $q_{\rm th}$. Throughout most of the envelope, $t_{\rm sc} \ll t_{\rm diff}$, indicating that the envelopes reside in the slow-diffusion, nearly adiabatic regime.}
\label{fig:lum_tcool_T80000}
\end{figure}

To assess the self-consistency of the adiabatic assumption, 
we calculate 
the thermal diffusion timescale towards the envelope boundary (which we assume to be $R_H$) as 

\begin{equation}
    t_{\rm diff}(R) = \tau(R) \dfrac{R_H- R}{c},
    \label{eqn:tdiff}
\end{equation}
assuming electron-scattering dominates the opacity, $\kappa=0.34$cm$^2$/g, with $\tau(R) = \int_R^{R_H} \kappa \rho dR$ being the outward optical depth. 
This timescale can be compared to the sound-crossing time 

\begin{equation}
    t_{\rm sc} ={\int_R^{R_H} dr/c_s(R)},
\end{equation}
which is typically much shorter than the diffusion time, $t_{\rm sc} \ll t_{\rm diff}$, 
throughout most of the envelope in our simulations 
(Figure~\ref{fig:lum_tcool_T80000}). This is similar to the local criterion $\tau > c/c_s$ \citep{Jiang2015}.
This ordering demonstrates that once formed, the envelopes must reside in the slow-diffusion regime, 
where radiation evolves nearly adiabatically coupled to the gas on dynamical (sound-crossing) timescales. 
In the slow-diffusion regime considered here, the diffusion time $t_{\rm diff}$ controls the thermal relaxation towards 
a quasi-steady state (e.g., Figure~\ref{fig:time_evo_T80000}). 
Because $t_{\rm diff}$ can be orders of magnitude longer than $t_{\rm sc}$, fully self-consistent three-dimensional RHD simulations, which are typically limited to only a few sound-crossing times, are computationally impractical for studying this regime.
By contrast, 
the simulations of \citet{Chen2025} probe a distinctly different parameter space, 
characterized by background densities lower by roughly two orders of magnitude. 
In that regime, 
radiative diffusion is rapid ($t_{\rm sc} \gtrsim t_{\rm diff}$) and plays a dynamically essential role, 
and the sound-crossing time $t_{\rm sc}$ sets the timescale for the system to approach thermal equilibrium, 
making full RHD simulations feasible and necessary. 

Nevertheless, even in the adiabatic regime, 
there may exist a very thin layer near the envelope boundary where $t_{\rm diff} \lesssim t_{\rm sc}$ and radiative cooling at this surface layer can be locally efficient. 
However, this is unlikely to significantly modify the shock-generated entropy 
that ultimately regulates the envelope mass and the onset of runaway accretion, 
because the accretion shock itself 
is embedded in the disk midplane.
In this sense, Equation~\ref{eqn:tdiff}, 
when integrated radially up to the \textit{envelope boundary} to define an approximately isotropic diffusion timescale, 
likely underestimates the true cooling time relevant for the midplane shock. 
In practice, radiation reaching the shock front must diffuse vertically toward the disk surface to cool, 
introducing an additional delay.

To test this, we restarted the simulation \texttt{D1e-8T8e4m50} from its quasi-steady state, 
and ran it for $\sim10$ additional orbits with implicit radiative transport \citep{Jiang2021}. 
The envelope shows no qualitative change in structure or mass over this interval.
Radiation transport being computationally expensive, we have not similarly extended the other models in Table~\ref{tab:parameters}.

\subsection{Varying background parameters}

\begin{figure}
    \centering
    \includegraphics[width=0.45\textwidth]{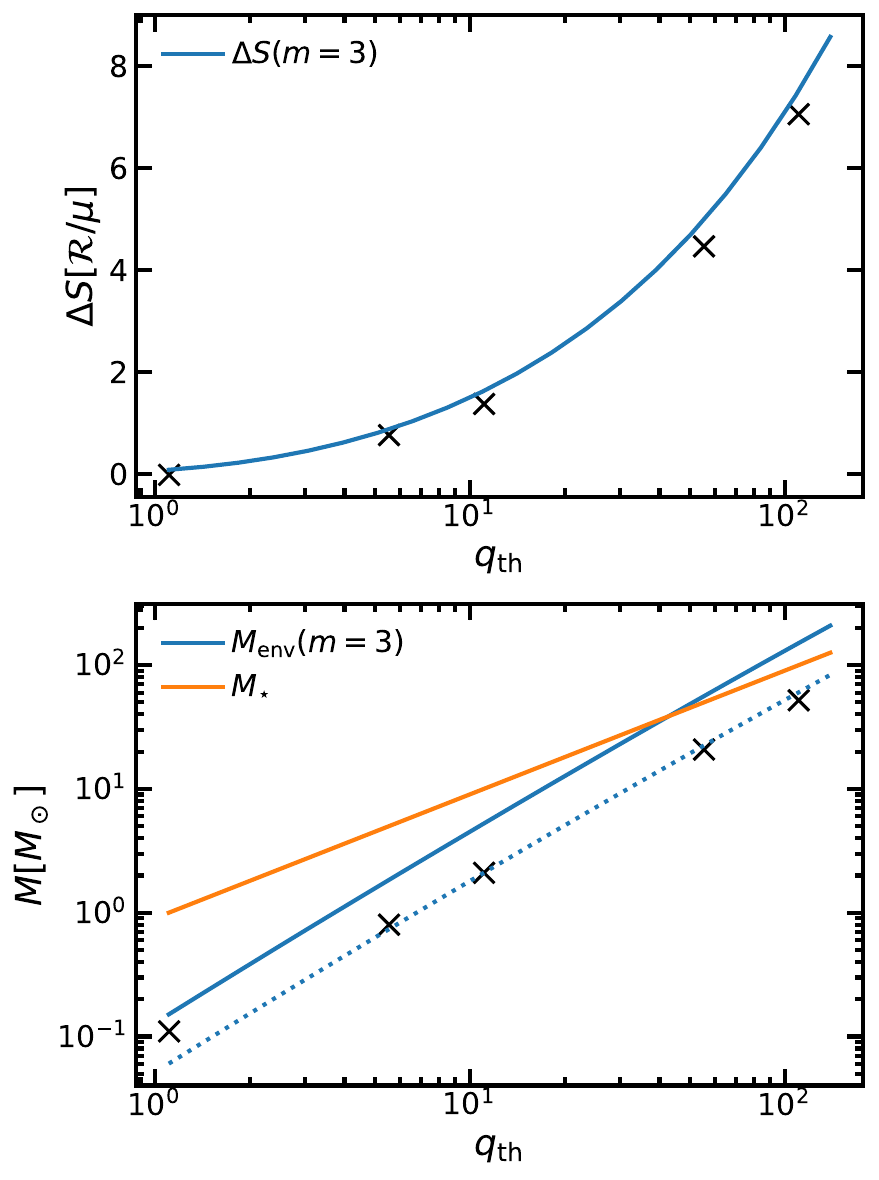}
    \caption{Similar to Figure \ref{fig:summary_T80000} but for $\rho_0 = 10^{-8},\mathrm{g,cm^{-3}}$ and $T_0 = 4\times10^4,\mathrm{K}$ simulations. }
\label{fig:summary_T40000}
\end{figure}

In Figure \ref{fig:summary_T40000} 
we present the calculation of 
$\Delta S$ and $M_{\rm env}$ for another set of simulations at $T_0 = 40000$K. 
The background disk midplane is colder with lower radiation to gas pressure ratio  $\Pi_0 = 0.12$ and lower entropy $S_0 = 0.48$, 
but the 
general trend for $\Delta S$ and $M_{\rm env}$ is similar to the $T_0 = 80000$K simulations. 
Note that due to much slower disk sound speed, the Bondi radii of $100 M_\odot$ 
stars are much larger than the scale height 
and even the simulation domain such that $q_{\rm th}$ reaches $ \approx 100$. 
However, 
in our superthermal cases 
the envelope limit is largely set by the Hill radius so the large nominal values of $R_B$ are not relevant.

\subsection{Self-gravitating simulations}
\label{sec:sg}

Once $M_{\rm env}$ exceeds $M_\star$, 
the inclusion of envelope self-gravity is expected to trigger runaway accretion \citep{Chen2024}. 
This transition should occur at sufficiently large $M_\star$ and/or high background density (see \S \ref{sec:general_runaway} for a more general discussion). 
To explore the outcome of this scenario, we perform additional runs \texttt{D1e-8T4e4m100\_sg} and \texttt{D3e-8T4e4m100\_sg}, 
where we include self-gravity of the gas within the simulation domain using the Poisson solver implemented in \texttt{Athena++} by \citet{Xu2021, Xu2025} based on spherical harmonics. 
In both cases, we first evolve the system without self-gravity until a quasi-steady envelope has been acquired. At this stage, run \texttt{D1e-8T4e4m100\_sg} is identical to \texttt{D1e-8T4e4m100}. 
We then restart the simulations with self-gravity source terms enabled in the momentum and energy equations at $t \simeq 15$ orbits, 
and continue the evolution to investigate whether the system relaxes to a new quasi-steady state or undergoes runaway accretion.

\begin{figure}
    \centering
    \includegraphics[width=0.43\textwidth]{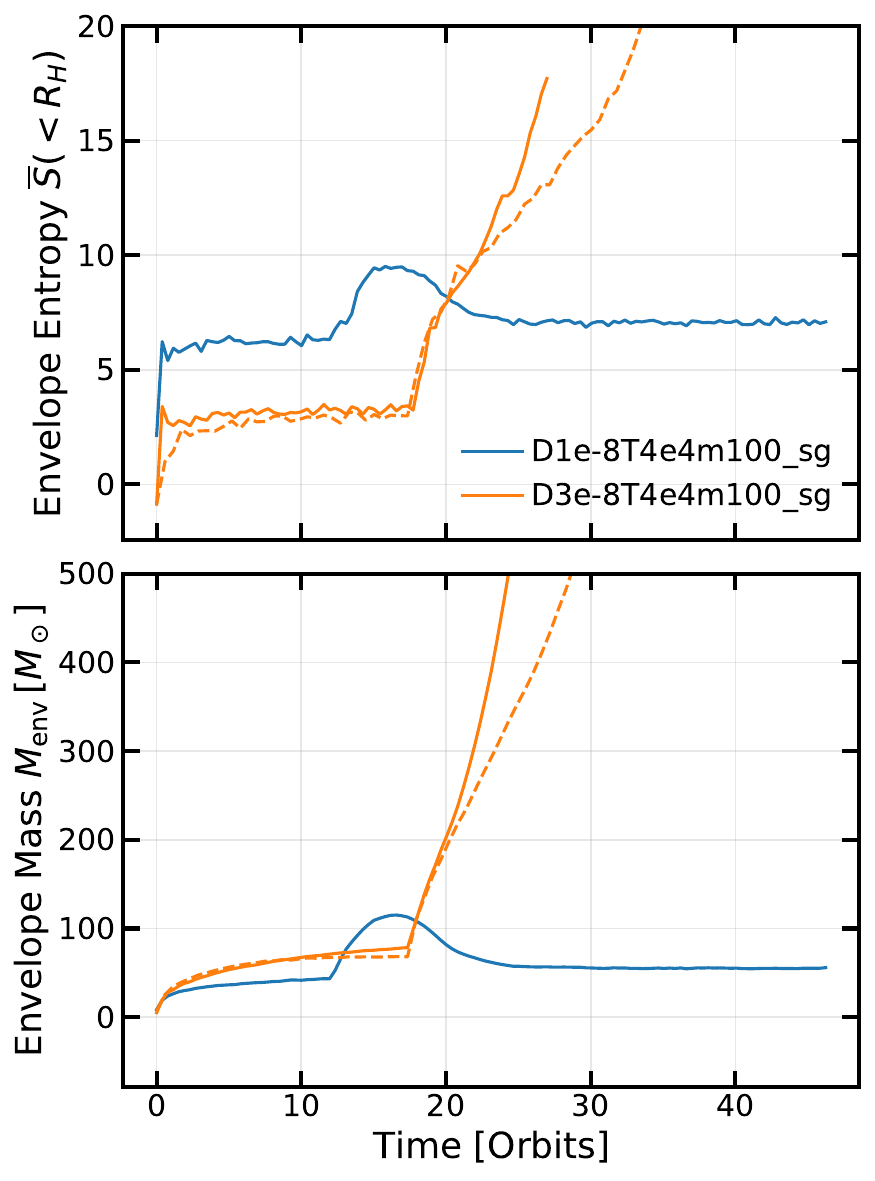}
    \caption{evolution of entropy and mass of envelope for simulations where we turn on self-gravitating midway. For run \texttt{D3e-8T4e4m100\_sg} no steady state is reached 
    and we need to stop the simulation when the Hill radius of the combined star+envelope mass reaches the radial extent of the simulation domain. The dashed line indicates a simulation with larger simulation domain which also reached a runaway stage. }
\label{fig:time_evo_sg}
\end{figure}

Figure \ref{fig:time_evo_sg} illustrates the impact of activating gas self-gravity on the evolution of the envelope entropy and mass. 
In both runs, turning on self-gravity produces an immediate, renewed phase of accretion, reflected by a rapid rise in the enclosed envelope mass and a concurrent increase in the mass-weighted entropy.
In run \texttt{D1e-8T4e4m100\_sg}, this initial boost is transient. After a period of adjustment, the system relaxes to a new quasi-steady state in which the final envelope mass exceeds that of the corresponding non-self-gravitating run \texttt{D1e-8T4e4m100} by 20\%. 
Despite the larger mass, equilibrium is maintained owing to enhanced thermal support associated with the elevated entropy profile, which counteracts further gravitational collapse. 
On the other hand, 
in run \texttt{D3e-8T4e4m100\_sg}, the inclusion of self-gravity drives sustained, runaway growth of the envelope mass. 
Although the entropy also increases following the onset of self-gravity, 
the resulting thermal support is insufficient to halt accretion, 
and the system does not settle into a quasi-steady state.

We note that our measurement of envelope mass and entropy within $R_H$ 
is only well-defined in the regime $M_{\rm env} \ll M_\star$,
where the Hill radius is set by the stellar mass alone. 
Once runaway accretion begins, however, 
the effective Hill radius of the system increases as $R_H^{'} = r_\bullet [( M_\star + M_{\rm env})/M_\bullet]^{1/3}$. 
As a result, the growing envelope within $\sim R_H^{'}$ with can approach radial extent of the simulation domain.
To verify that the observed runaway behavior is not an artifact of the finite domain size, 
we performed an additional convergence test using a simulation otherwise identical to \texttt{D3e-8T4e4m100\_sg}, 
but with the radial domain extended to 3000$R_\odot$ (dashed lines in Figure \ref{fig:time_evo_sg}). 
This simulation also exhibits qualitatively similar runaway accretion 
after self-gravity is enabled, demonstrating that the runaway behavior is physical.

\subsection{Rotation and Angular Momentum Transport}
\label{sec:rotation}

It is well known that conservation of vortensity in a corotating frame with orbital frequency $\Omega$,

\begin{equation}
    \zeta := \dfrac{(\nabla \times \vec{v})\cdot \hat{z}+2\Omega}{\Sigma}\,,
\end{equation}
implies that, insofar as the flow is two-dimensional and isentropic, 
gas drawn from a $\Sigma_0$ density background with Keplerian shear, carrying $\zeta \approx 0.5\Omega/\Sigma_0$, 
should acquire 
strong prograde rotation as it is compressed to higher surface densities $\Sigma \gg \Sigma_0$ near an embedded companion. 
This tendency naturally favors the formation of a rotationally supported, disk-like structure, 
largely independent of the equation of state \citep{Papaloizou1989,Korycansky1996}. 
In contrast, 
in our quasi-steady envelopes we consistently measure azimuthal velocities of only $\sim10-20\%$ of the local Keplerian velocity (e.g., Figure~\ref{fig:T8e4_rho1e-8_qth1_radial}), 
indicating that rotation provides only weak support. 
This implies that any 
sustained inward advection of angular momentum must be balanced by an outward transport mechanism. 
In the quasi-steady state, this poses no difficulty because we find that, consistent with the nearly vanishing net accretion rate ($\dot{M}\sim 0$), the net advected angular momentum flux is also negligible after averaging over the simultaneous inflowing and outflowing gas, despite the envelope being only marginally bound.
By contrast, 
once runaway begins, 
the non-zero net mass inflow is accompanied by a substantial inward angular momentum flux, 
which necessarily requires outward transport through non-axisymmetric and/or time-dependent stresses to sustain continued accretion. 
Spiral shocks are a plausible contributor to this transport \citep{Chen2025}, but convection, turbulence, and other 3D motions may also play important roles. 
A detailed decomposition of these contributions, however, is beyond the scope of the present work.

\section{Discussions and Summary}
\label{sec:discussions}

\subsection{General Condition for Runaway}
\label{sec:general_runaway}
Using a combination of numerical simulations and semi-analytical arguments, 
we have shown phenomenologically that runaway accretion sets in once the envelope mass becomes comparable to or exceeds the stellar mass, 
$M_{\rm env} \gtrsim M_\star$, which occurs at sufficiently large $q_{\rm th}$ and low Toomre $Q$. 
We can, nevertheless, provide a more fundamental derivation and interpretation of this criterion from first principles. 
Following \citet{Chen2024}, 
we can calculate that in the adiabatic and non-self-gravitating limit, accretion from an idealized isotropic and non-rotating background
builds an isentropic envelope 
within the stellar Bondi radius, $R_{\rm env}\sim R_B$, with a total mass of approximately

\begin{equation}
M_{\rm env,iso} \approx 4\pi \rho_0 R_B^3 \ln (R_{\rm env}/R_{\rm in})
=  4\pi f (G M_\star)^3 \frac{\rho_0}{ c_{s,0}^6} ,
\label{eqn:Menviso}
\end{equation}
or at least in the radiation-pressure-dominated regime where $\gamma \simeq 4/3$ and the envelope mass is dominated by its outer layers. 
Here $f = \ln (R_{\rm env}/R_{\rm in}) \sim \mathcal{O}(1)$ is a logarithmic factor.
On the other hand, for radiation-pressure-dominated stars, the $n=3$ polytropic relation gives the stellar mass for an Eddington model

\begin{equation}
M_*^2 \approx \frac{1}{(0.3639 G)^3}\left(\frac{3}{a}\right)\left(\frac{\mathcal{R}}{\mu_{\star} }\right)^4 \frac{1-\beta_*}{\beta_*^4}\,,
\label{eqn:Eddington}
\end{equation}
where the gas-pressure fraction $\beta_\star = (1+\Pi_\star)^{-1} \sim \Pi_\star^{-1}$ when $\Pi_\star\gg 1$. 
Using this substitution, Equation \eqref{eqn:Menviso} can be recast as [omitting $f\sim O(1)$]

\begin{equation}
    M_{\mathrm{env,iso}} / M_* \sim   \left(\frac{\mu}{\mu_*}\right)^4 \times \left(\frac{\Pi_\star}{\Pi_0}\right)^4.
    \label{eqn:Menviso_entropy}
\end{equation}

When the background molecular weight $\mu\approx\mu_\star$, 
this formula implies 
$M_{\mathrm{env,iso}} / M_*>1$ if the background material has a lower radiation entropy (i.e., lower $\Pi$) than the stellar interior.
In other words, 
once the envelope mass exceeds the stellar mass, the characteristic entropy associated with an $n=3$ polytrope of mass $M_\star$ is higher than that of the material accreting into the envelope. 
The lower-entropy inflowing gas therefore tends to sink toward the center via convective mixing. 
In the presence of sustained mass inflow, maintaining a stably stratified hydrostatic configuration becomes increasingly difficult, 
naturally leading to runaway growth.

The discussion above pertains to accretion from an asymptotically stationary, isotropic environment, and neglects shocks. 
This may be appropriate in the subthermal regime ($q_\mathrm{th}<1$) when both the Bondi radius and the Hill radius are smaller than the disk scale height (cf. \S \ref{sec:setup}).
But how does this picture connect to the runaway criterion in a stratified disk environment? Using $H=c_{s,0}/\Omega$ and $\Omega=\sqrt{G M_\bullet/r^{3}}$, 
Equation~\eqref{eqn:Menviso} can be rewritten in terms of the Toomre parameter as

\begin{equation}\label{eq:MQq}
\frac{M_{\rm env,iso}}{M_\star} \sim \frac{2 q_{\rm th}^2}{Q}.
\end{equation}

This expression highlights an apparent contradiction. 
In an idealized isotropic medium, 
dynamical runaway is expected once $M_{\rm env,iso}/M_\star>1$. 
However, in a disk environment that is non-self-gravitating ($Q > 1$), eq.~\eqref{eq:MQq} states that $M_{\rm env,iso}/M_\star>1$ requires $q_{\rm th}\gtrsim 1$. 
In this regime, tidal truncation becomes important and the assumption that $R_{\rm env}\sim R_B<R_H$ fails. 
Rather, the envelope size is limited by the Hill radius, $R_{\rm env}\sim R_H < R_B$.
As a result, the actual envelope mass is capped below the isotropic estimate \eqref{eqn:Menviso_entropy} 
and the criterion for dynamical runaway becomes more restrictive.

Guided by simulations in this paper, 
it is straightforward to extend the above calculations to the high-$q_{\rm th}$ regime 
relevant for a more realistic, tidally limited envelope. 
Although complexities remain about the partition of radiation and gas pressure in supporting the envelope in the $q_{\rm th} \sim 1$ regime (See \S \ref{sec:entropy_mass}), 
we can infer that as $q_{\rm th}$ and the pre-shock Mach number increases, 
the post-shock envelope becomes increasingly radiation supported, 
approaching an effective adiabatic index $\gamma_{\rm eff}\simeq4/3$. 
In this limit, 
Equation \eqref{eqn:compression_rho} yields a fixed compression ratio

\begin{equation}
    \rho'/\rho_0 \approx \dfrac{\gamma_{\rm eff}+1}{\gamma_{\rm eff}-1} = 7,
\end{equation}
which means that at large $q_{\rm th}$ the envelope relaxes to an approximately $n=3$ polytropic structure and $ M_{\rm env} /M_{\star}$ can at most grow as

\begin{equation}
    M_{\rm env} \approx 4\pi f \rho' R_H^3  \sim 3.5 M_\star/Q.
    \label{eqn:Menv}
\end{equation}

This means the disk background needs to be nearly self-gravitating ($Q \lesssim 3.5$) for runaway accretion to be possible at high $q_{\rm th}$, as shown by the local simulations presented in this paper. 
In the disk of a bright AGN,
the outer regions may maintain $Q \sim 1$ \citep{Goodman2003}. 
Within this self-gravitating zone, 
the midplane density increases toward smaller radii and reaching a maximum near the transition to a standard, accretion-powered disk at the minimum self-gravitating radius $r_{\rm sg}$. 
This radius could be most favorable location for runaway accretion: 
it combines the highest ambient density with large optical depth that can be achieved in a global multi-zone AGN disk model \citep{Sirko2003}, 
making accretion most likely to proceed in the adiabatic regime (also see \S \ref{sec:summary}). 


We can interpolate between the $q_{\rm th} \ll 1$ and $q_{\rm th} \gg 1$ limits and write

\begin{equation}
\frac{M_{\rm env}}{M_\star}
\sim \dfrac 1Q\min\left(3.5,2q_{\rm th}^2\right).
\end{equation}

In summary, 
we predict runaway only if this ratio reaches unity. 
In practice, only the high-$q_{\rm th}$ branch is relevant for realistic stellar growth in AGN disks. 
In that regime, the criterion reduces to requiring the background disk be marginally self-gravitating.
The logarithmic factor $f>1$ modestly enlarges the parameter space for runaway, but does not qualitatively alter this criterion.

Finally, in close analogy to Equation~\eqref{eqn:Menviso_entropy} which provides an entropy-based interpretation of Equation~\eqref{eqn:Menv}, we can also formulate the runaway criterion in the high $q_{\rm th}$ limit in terms of a radiation entropy ratio.
For $q_{\rm th} \gg 1$, Equation~\eqref{eqn:compression_P} yields
\begin{equation}\label{eq:Ppvsq}
\frac{P'}{P_0} \approx (3q_{\rm th})^{2/3},
\qquad
\frac{\Pi'}{\Pi_0}
\approx \left[\frac{(3q_{\rm th})^{2/3}}{7^{4/3}}\right]^{3/4}
\sim q_{\rm th}^{1/2},
\end{equation}
indicating that the post-shock material entering the envelope has substantially enhanced radiation entropy. The corresponding ``runaway by infalling low-entropy material” criterion can then be written as 

\begin{equation}
\frac{M_{\rm env}}{M_\star}
\sim \left(\frac{\mu}{\mu_\star}\right)^4
\left(\frac{\Pi_\star}{\Pi'}\right)^4
\sim
\frac{M_{\rm env,iso}}{M_\star} \min(1,q_\mathrm{th}^{-2}).
\end{equation}

The physical meaning of this expression is straightforward: 
at $q_{\rm th}\gg1$, 
the tidally-induced accretion shock significantly increases the entropy of the material flowing into the envelope, 
by a factor scaling as $\sim q_{\rm th}^2$. 
As a result, it becomes increasingly difficult for the characteristic radiation entropy of the star to exceed that of the inflowing gas, 
which is necessary to trigger runaway accretion. 
In this sense, 
the underlying physics mirrors that of the isotropic case, 
only with the shock-modified entropy replacing the background entropy.

\subsection{Relevance to Planetary Runaway}

The mass or entropy ratio criterions derived above invites comparison with the classical runaway accretion of planetary envelopes \citep{Mizuno1980,Stevenson1982,Pollack1996,Piso2014,LeeChiangOrmel2014,Ali-Dib2020,ChenLi2020}. 
In both cases, 
runaway occurs when an initially hydrostatic envelope loses sufficient thermal support and contracts under its own gravity. 
Nevertheless, the thermodynamic regimes underlying the two processes are quite different.
In the planetary scenario, contraction is always regulated by radiative cooling. 
The entropy of the rocky core is substantially lower than that of the surrounding protoplanetary disk, 
such that convective recycling, 
atmospheric replenishment, 
and advection of disk material can only inject energy into the envelope, 
suppressing accretion \citep{Ormel2015,Lambrechts2017,Kurokawa2018,Bailey2024}.
Consequently, runaway proceeds only if radiative diffusion can remove thermal energy sufficiently rapidly, and is not expected to occur in an adiabatic scenario. 

The picture is qualitatively different when the entropy ordering is reversed. 
As discussed in \S \ref{sec:general_runaway}, 
adiabatic runaway ultimately requires the characteristic entropy of the stellar interior to exceed that of the surrounding AGN disk and even the post-shock envelope. 
Under these conditions, convective or advective transport no longer replenishes the envelope with thermal energy. 
Instead, these processes could transport energy outward, 
allowing the envelope to lose (or effectively, unable to maintain) pressure support in the absence of radiative diffusion. 
In this sense, the onset of runaway no longer requires radiative cooling, although radiative diffusion, 
if included, 
would simply provide an additional energy-loss channel and further promote contraction. 
Our adiabatic calculations are therefore conservative with respect to the onset of runaway.

In summary, although both instabilities ultimately involve the loss of thermal support and the growth of envelope self-gravity, 
the mechanism identified here should be regarded as distinct from classical planetary runaway. 

\subsection{Final Outcome of Runaway}
\label{sec:gap}
With no steady hydrostatic equilibrium to halt accretion, 
we expect the star to grow to the isolation mass $M_{\rm iso}$ \citep{Lissauer1993,GoodmanTan2004}, 
which contains the total disk mass within an annulus of width $R_H(M_{\rm iso})$: 

\begin{equation}
    M_{\rm iso} = \dfrac{M_{\rm disk}^{3/2}}{ M_\bullet^{1/2}},\quad   M_{\rm disk}:= 4\pi \rho_0 H r^2
\end{equation}

Expressed in terms of the disk aspect ratio $h \equiv H/r$, the isolation mass becomes $  M_{\rm iso}\approx (h/Q)^{3/2} M_\bullet$. 
For superthermal companions, 
the dynamical (runaway) accretion rate in the ``Hills regime" scales with $q^{2/3}$, where $q$ (not to be confused with $q_\mathrm{th}$) is the mass ratio of the accreting satellite to that of the central mass \citep{Li2023,Choksi2023}. 
Adopting this rate, the timescale to reach the isolation mass is \citep[][see their Equation~36]{GoodmanTan2004}
\begin{equation}
t_{\rm iso}
\approx \frac{2\pi}{\Omega}
\left(\frac{M_\bullet}{M_{\rm disk}}\right)^{1/2}
\approx \frac{2\pi}{\Omega}
\left(\frac{Q}{h}\right)^{1/2}.
\end{equation}
For self-gravitating disks with $Q\sim 1$ and $h \gtrsim 10^{-3}\mbox{-}10^{-2}$ \citep{Sirko2003,Thompson2005,ChenLiu2026}, 
this corresponds to a relatively short growth timescale of $\sim 10\mbox{-}100$ orbital periods.

However, this estimate implicitly assumes that the characteristic density of the mass supply,
$\rho_0$, 
remains approximately constant over the isolation timescale $t_{\rm iso}$. In practice, in the large-$q$ limit, gap-opening may occur on the synodic timescale:

\begin{equation}
t_{\rm syn} \approx \frac{2\pi}{\Omega} q^{-1/3}.
\end{equation}

This reflects the requirement that disk material must pass through the companion’s vicinity at least $\mathcal{O}(1)$ times in order to be significantly deflected and exchange angular momentum, 
effectively experiencing the Lindblad torque.
Since $t_{\rm syn}$ decreases with increasing $q$, 
for a supermassive star growing toward the isolation mass the minimum deflection/depletion timescale satisfies
\begin{equation}
t_{\rm syn}\left(q=\frac{M_{\rm iso}}{M_\bullet}\right)
\approx t_{\rm iso}.
\end{equation}
This implies that, 
even in the inviscid limit, the depletion timescale is likely comparable to or longer than the accretion timescale, 
allowing the star to approach the isolation mass before local disk material is significantly depleted.
The inclusion of viscosity, 
parameterized by $\alpha$ \citep{SS1973}, 
introduces an additional viscous timescale $t_{\rm vis} \sim \alpha^{-1}$ orbital periods. 
When $t_{\rm vis} < t_{\rm syn}$, 
viscous diffusion can partially refill the gap and further delay local depletion of disk material.
The overall influence of these 
global effects may impose quantitative changes on the final isolation mass $M_{\rm iso}$. 
Assessing these effects requires global simulations that capture full azimuthal structures of the SMBH accretion disk.

\subsection{Summary}
\label{sec:summary}
In this paper we studied the properties of the circumstellar envelope 
acquired by massive stars embedded in AGN disks in the adiabatic or 
slow-diffusion limit. 
In the transsonic regime ($q_{\rm th}\sim 1$), 
the entropy of the envelope and the boundary density smoothly match the ambient disk conditions. 
However, 
in the superthermal regime $q_{\rm th} = q/h^3 >1$, or equivalently when the stellar mass $M_\star$ is above the thermal mass $M_{\rm th} = c_{s,0}^3/G\Omega$, 
the envelope is coupled to the background through a strong shock. 
This shock substantially enhances both the density and entropy of the envelope, 
yielding an envelope whose mass scales as $M_{\rm env} \sim M_\star/Q$ 
in the limit $q_{\rm th} \gg 1$, and is likely to be radiation-pressure-dominated due to high post-shock entropy. 
In marginally self-gravitating disks ($Q\sim 1$), 
this can lead to runaway growth of the envelope once self-gravity is included in our simulations. 
We expect this runaway to terminate near the isolation mass, $\sim 10^4$ -$10^5 M_\odot$,  
when the Hill radius of the system has been evacuated. 
This outcome contrasts sharply with the more commonly discussed picture 
in which disk fragmentation produces multiple stars that grow slowly and saturate at moderate masses \citep{Levin+Beloborodov2003,Levin2003}.
This dichotomy is mostly controlled by the optical depth or radiative diffusion timescale. 
When the optical depth is sufficiently 
large for accretion to proceed in the slow-diffusion regime, 
runaway growth becomes possible; 
Otherwise,
stellar accretion luminosity and associated efficient radiative feedback are 
sufficient to regulate stellar growth and prevent runaway \citep{Chen2025,Xu2025}. 
The former scenario 
likely applies to realistic environments near the minimum self-gravitating radius of an AGN disk  \citep{Goodman2003,GoodmanTan2004,Chen2024}. 

After supermassive stars form in the vicinity of massive black holes, they may give rise to distinctive observational signatures, 
such as pair-instability supernovae, stellar mergers, or extreme-mass-ratio inspiral events \citep{GoodmanTan2004}. 
This may also provide a channel for forming intermediate mass black holes in AGN disks \citep{McKernan2012}.
Exploring the detectability and observational consequences of these outcomes 
is an important direction for subsequent studies.

\section*{Acknowledgements}
We thank Wenrui Xu for providing us with the spherical-polar self-gravity solver and Philippe Yao for the Eddington equation of state module. We thank Douglas Lin, Zhaohuan Zhu and Adam Burrows for helpful discussions and the anonymous referee for constructive comments.



\bibliography{sample631}{}

\begin{thebibliography}{}
\expandafter\ifx\csname natexlab\endcsname\relax\def\natexlab#1{#1}\fi
\providecommand{\url}[1]{\href{#1}{#1}}
\providecommand{\dodoi}[1]{doi:~\href{http://doi.org/#1}{\nolinkurl{#1}}}
\providecommand{\doeprint}[1]{\href{http://ascl.net/#1}{\nolinkurl{http://ascl.net/#1}}}
\providecommand{\doarXiv}[1]{\href{https://arxiv.org/abs/#1}{\nolinkurl{https://arxiv.org/abs/#1}}}

\bibitem[{{Ali-Dib} {et~al.}(2020){Ali-Dib}, {Cumming}, \& {Lin}}]{Ali-Dib2020}
{Ali-Dib}, M., {Cumming}, A., \& {Lin}, D. N.~C. 2020, \mnras, 494, 2440,
  \dodoi{10.1093/mnras/staa914}

\bibitem[{{Ali-Dib} \& {Lin}(2023)}]{AliDib2023}
{Ali-Dib}, M., \& {Lin}, D. N.~C. 2023, \mnras, \dodoi{10.1093/mnras/stad2774}

\bibitem[{{Artymowicz} {et~al.}(1993){Artymowicz}, {Lin}, \&
  {Wampler}}]{Artymowicz1993}
{Artymowicz}, P., {Lin}, D.~N.~C., \& {Wampler}, E.~J. 1993, \apj, 409, 592,
  \dodoi{10.1086/172690}

\bibitem[{{Bailey} \& {Zhu}(2024)}]{Bailey2024}
{Bailey}, A.~P., \& {Zhu}, Z. 2024, \mnras, 534, 2953,
  \dodoi{10.1093/mnras/stae2250}

\bibitem[{{B{\'e}thune} \& {Rafikov}(2019)}]{Bethune2019}
{B{\'e}thune}, W., \& {Rafikov}, R.~R. 2019, \mnras, 487, 2319,
  \dodoi{10.1093/mnras/stz1427}

\bibitem[{{Cantiello} {et~al.}(2021){Cantiello}, {Jermyn}, \&
  {Lin}}]{Cantiello2021}
{Cantiello}, M., {Jermyn}, A.~S., \& {Lin}, D. N.~C. 2021, \apj, 910, 94,
  \dodoi{10.3847/1538-4357/abdf4f}

\bibitem[{{Chen} {et~al.}(2022){Chen}, {Bailey}, {Stone}, \& {Zhu}}]{Chen2022}
{Chen}, Y.-X., {Bailey}, A., {Stone}, J., \& {Zhu}, Z. 2022, \apjl, 939, L23,
  \dodoi{10.3847/2041-8213/ac9b3e}

\bibitem[{{Chen} {et~al.}(2025){Chen}, {Jiang}, \& {Goodman}}]{Chen2025}
{Chen}, Y.-X., {Jiang}, Y.-F., \& {Goodman}, J. 2025, \apj, 987, 188,
  \dodoi{10.3847/1538-4357/addd0a}

\bibitem[{{Chen} {et~al.}(2024){Chen}, {Jiang}, {Goodman}, \& {Lin}}]{Chen2024}
{Chen}, Y.-X., {Jiang}, Y.-F., {Goodman}, J., \& {Lin}, D. N.~C. 2024, \apj,
  974, 106, \dodoi{10.3847/1538-4357/ad6dd4}

\bibitem[{{Chen} {et~al.}(2023){Chen}, {Jiang}, {Goodman}, \&
  {Ostriker}}]{Chen2023}
{Chen}, Y.-X., {Jiang}, Y.-F., {Goodman}, J., \& {Ostriker}, E.~C. 2023, \apj,
  948, 120, \dodoi{10.3847/1538-4357/acc023}

\bibitem[{{Chen} {et~al.}(2020){Chen}, {Li}, {Li}, \& {Lin}}]{ChenLi2020}
{Chen}, Y.-X., {Li}, Y.-P., {Li}, H., \& {Lin}, D. N.~C. 2020, \apj, 896, 135,
  \dodoi{10.3847/1538-4357/ab9604}

\bibitem[{{Chen} {et~al.}(2026){Chen}, {Liu}, {Li}, {Wang}, {Ma}, {Jiang},
  {Greene}, {Quataert}, \& {Goodman}}]{ChenLiu2026}
{Chen}, Y.-X., {Liu}, H., {Li}, R., {et~al.} 2026, \apjl, 1003, L46,
  \dodoi{10.3847/2041-8213/ae6b69}

\bibitem[{{Choksi} {et~al.}(2023){Choksi}, {Chiang}, {Fung}, \&
  {Zhu}}]{Choksi2023}
{Choksi}, N., {Chiang}, E., {Fung}, J., \& {Zhu}, Z. 2023, \mnras, 525, 2806,
  \dodoi{10.1093/mnras/stad2269}

\bibitem[{{Coleman}(2020)}]{Coleman2020}
{Coleman}, M. S.~B. 2020, \apjs, 248, 7, \dodoi{10.3847/1538-4365/ab82ff}

\bibitem[{{Collin} \& {Zahn}(1999)}]{Collin+Zahn1999}
{Collin}, S., \& {Zahn}, J.-P. 1999, \apss, 265, 501,
  \dodoi{10.1023/A:1002191506811}

\bibitem[{{Dittmann} {et~al.}(2021){Dittmann}, {Cantiello}, \&
  {Jermyn}}]{Dittmann2021}
{Dittmann}, A.~J., {Cantiello}, M., \& {Jermyn}, A.~S. 2021, \apj, 916, 48,
  \dodoi{10.3847/1538-4357/ac042c}

\bibitem[{{Epstein-Martin} {et~al.}(2024){Epstein-Martin}, {Tagawa}, {Haiman},
  \& {Perna}}]{Epstein-Martin2024}
{Epstein-Martin}, M., {Tagawa}, H., {Haiman}, Z., \& {Perna}, R. 2024, arXiv
  e-prints, arXiv:2405.09380, \dodoi{10.48550/arXiv.2405.09380}

\bibitem[{{Fabj} {et~al.}(2025){Fabj}, {Dittmann}, {Cantiello}, {Perna}, \&
  {Samsing}}]{Fabj2025}
{Fabj}, G., {Dittmann}, A.~J., {Cantiello}, M., {Perna}, R., \& {Samsing}, J.
  2025, \apj, 981, 16, \dodoi{10.3847/1538-4357/ada896}

\bibitem[{{Floris} {et~al.}(2024){Floris}, {Marziani}, {Panda}, {Sniegowska},
  {D'Onofrio}, {Deconto-Machado}, {Del Olmo}, \& {Czerny}}]{Floris2024}
{Floris}, A., {Marziani}, P., {Panda}, S., {et~al.} 2024, arXiv e-prints,
  arXiv:2405.04456, \dodoi{10.48550/arXiv.2405.04456}

\bibitem[{{Fryer} {et~al.}(2025){Fryer}, {Huang}, {Ali-Dib}, {Andrews}, {Xu},
  \& {Lin}}]{Fryer2025}
{Fryer}, C.~L., {Huang}, J., {Ali-Dib}, M., {et~al.} 2025, \mnras, 537, 1556,
  \dodoi{10.1093/mnras/staf130}

\bibitem[{{Fung} {et~al.}(2019){Fung}, {Zhu}, \& {Chiang}}]{Fung2019}
{Fung}, J., {Zhu}, Z., \& {Chiang}, E. 2019, \apj, 887, 152,
  \dodoi{10.3847/1538-4357/ab53da}

\bibitem[{{Goodman}(2003)}]{Goodman2003}
{Goodman}, J. 2003, \mnras, 339, 937, \dodoi{10.1046/j.1365-8711.2003.06241.x}

\bibitem[{{Goodman} \& {Rafikov}(2001)}]{Goodman2001}
{Goodman}, J., \& {Rafikov}, R.~R. 2001, \apj, 552, 793, \dodoi{10.1086/320572}

\bibitem[{{Goodman} \& {Tan}(2004)}]{GoodmanTan2004}
{Goodman}, J., \& {Tan}, J.~C. 2004, \apj, 608, 108, \dodoi{10.1086/386360}

\bibitem[{{Graham} {et~al.}(2025){Graham}, {McKernan}, {Ford}, {Stern},
  {Cantiello}, {Drake}, {Ding}, {Kasliwal}, {Koss}, {Margutti}, {Rose},
  {Somalwar}, {Wiseman}, {Djorgovski}, {Veres}, {Bellm}, {Chen}, {Groom},
  {Kulkarni}, \& {Mahabal}}]{Graham2025}
{Graham}, M.~J., {McKernan}, B., {Ford}, K.~E.~S., {et~al.} 2025, Nature
  Astronomy, \dodoi{10.1038/s41550-025-02699-0}

\bibitem[{{Hamann} \& {Ferland}(1999)}]{Hamann1999}
{Hamann}, F., \& {Ferland}, G. 1999, \araa, 37, 487,
  \dodoi{10.1146/annurev.astro.37.1.487}

\bibitem[{{Huang} {et~al.}(2023){Huang}, {Lin}, \& {Shields}}]{Huang2023}
{Huang}, J., {Lin}, D. N.~C., \& {Shields}, G. 2023, \mnras, 525, 5702,
  \dodoi{10.1093/mnras/stad2642}

\bibitem[{{Jiang} \& {Pan}(2025)}]{JiangPan2025}
{Jiang}, N., \& {Pan}, Z. 2025, \apjl, 983, L18,
  \dodoi{10.3847/2041-8213/adc456}

\bibitem[{{Jiang}(2021)}]{Jiang2021}
{Jiang}, Y.-F. 2021, \apjs, 253, 49, \dodoi{10.3847/1538-4365/abe303}

\bibitem[{{Jiang} {et~al.}(2015){Jiang}, {Cantiello}, {Bildsten}, {Quataert},
  \& {Blaes}}]{Jiang2015}
{Jiang}, Y.-F., {Cantiello}, M., {Bildsten}, L., {Quataert}, E., \& {Blaes}, O.
  2015, \apj, 813, 74, \dodoi{10.1088/0004-637X/813/1/74}

\bibitem[{{Jiang} \& {Goodman}(2011)}]{Jiang11}
{Jiang}, Y.-F., \& {Goodman}, J. 2011, \apj, 730, 45,
  \dodoi{10.1088/0004-637X/730/1/45}

\bibitem[{{Korycansky} \& {Papaloizou}(1996)}]{Korycansky1996}
{Korycansky}, D.~G., \& {Papaloizou}, J.~C.~B. 1996, \apjs, 105, 181,
  \dodoi{10.1086/192311}

\bibitem[{{Kurokawa} \& {Tanigawa}(2018)}]{Kurokawa2018}
{Kurokawa}, H., \& {Tanigawa}, T. 2018, \mnras, 479, 635,
  \dodoi{10.1093/mnras/sty1498}

\bibitem[{{Lai} {et~al.}(2022){Lai}, {Bian}, {Onken}, {Wolf}, {Mazzucchelli},
  {Ba{\~n}ados}, {Bischetti}, {Bosman}, {Becker}, {Cupani}, {D'Odorico},
  {Eilers}, {Fan}, {Farina}, {Onoue}, {Schindler}, {Walter}, {Wang}, {Yang}, \&
  {Zhu}}]{Lai+2022}
{Lai}, S., {Bian}, F., {Onken}, C.~A., {et~al.} 2022, \mnras, 513, 1801,
  \dodoi{10.1093/mnras/stac1001}

\bibitem[{{Lambrechts} \& {Lega}(2017)}]{Lambrechts2017}
{Lambrechts}, M., \& {Lega}, E. 2017, \aap, 606, A146,
  \dodoi{10.1051/0004-6361/201731014}

\bibitem[{{Lee} {et~al.}(2014){Lee}, {Chiang}, \& {Ormel}}]{LeeChiangOrmel2014}
{Lee}, E.~J., {Chiang}, E., \& {Ormel}, C.~W. 2014, \apj, 797, 95,
  \dodoi{10.1088/0004-637X/797/2/95}

\bibitem[{{Levin}(2003)}]{Levin2003}
{Levin}, Y. 2003, arXiv e-prints, astro,
  \dodoi{10.48550/arXiv.astro-ph/0307084}

\bibitem[{{Levin} \& {Beloborodov}(2003)}]{Levin+Beloborodov2003}
{Levin}, Y., \& {Beloborodov}, A.~M. 2003, \apjl, 590, L33,
  \dodoi{10.1086/376675}

\bibitem[{{Li} {et~al.}(2023){Li}, {Chen}, \& {Lin}}]{Li2023}
{Li}, Y.-P., {Chen}, Y.-X., \& {Lin}, D. N.~C. 2023, \mnras, 526, 5346,
  \dodoi{10.1093/mnras/stad3049}

\bibitem[{{Li} {et~al.}(2022){Li}, {Chen}, {Lin}, \& {Wang}}]{Li2022}
{Li}, Y.-P., {Chen}, Y.-X., {Lin}, D. N.~C., \& {Wang}, Z. 2022, \apjl, 928,
  L1, \dodoi{10.3847/2041-8213/ac5b61}

\bibitem[{{Li} {et~al.}(2021){Li}, {Dempsey}, {Li}, {Li}, \& {Li}}]{Li2021}
{Li}, Y.-P., {Dempsey}, A.~M., {Li}, S., {Li}, H., \& {Li}, J. 2021, \apj, 911,
  124, \dodoi{10.3847/1538-4357/abed48}

\bibitem[{{Linial} \& {Metzger}(2023)}]{Linial2023}
{Linial}, I., \& {Metzger}, B.~D. 2023, arXiv e-prints, arXiv:2303.16231,
  \dodoi{10.48550/arXiv.2303.16231}

\bibitem[{{Lissauer}(1993)}]{Lissauer1993}
{Lissauer}, J.~J. 1993, \araa, 31, 129,
  \dodoi{10.1146/annurev.aa.31.090193.001021}

\bibitem[{{Lynden-Bell}(1969)}]{Lyndenbell1969}
{Lynden-Bell}, D. 1969, \nat, 223, 690, \dodoi{10.1038/223690a0}

\bibitem[{{MacLeod} \& {Lin}(2020)}]{MacLeod2020}
{MacLeod}, M., \& {Lin}, D. N.~C. 2020, \apj, 889, 94,
  \dodoi{10.3847/1538-4357/ab64db}

\bibitem[{{McKernan} {et~al.}(2014){McKernan}, {Ford}, {Kocsis}, {Lyra}, \&
  {Winter}}]{McKernan2014}
{McKernan}, B., {Ford}, K.~E.~S., {Kocsis}, B., {Lyra}, W., \& {Winter}, L.~M.
  2014, \mnras, 441, 900, \dodoi{10.1093/mnras/stu553}

\bibitem[{{McKernan} {et~al.}(2012){McKernan}, {Ford}, {Lyra}, \&
  {Perets}}]{McKernan2012}
{McKernan}, B., {Ford}, K.~E.~S., {Lyra}, W., \& {Perets}, H.~B. 2012, \mnras,
  425, 460, \dodoi{10.1111/j.1365-2966.2012.21486.x}

\bibitem[{{Mihalas} \& {Mihalas}(1984)}]{Mihilas1984}
{Mihalas}, D., \& {Mihalas}, B.~W. 1984, {Foundations of radiation
  hydrodynamics}

\bibitem[{{Mizuno}(1980)}]{Mizuno1980}
{Mizuno}, H. 1980, Progress of Theoretical Physics, 64, 544,
  \dodoi{10.1143/PTP.64.544}

\bibitem[{{Nagao} {et~al.}(2006){Nagao}, {Maiolino}, \& {Marconi}}]{Nagao2006}
{Nagao}, T., {Maiolino}, R., \& {Marconi}, A. 2006, \aap, 459, 85,
  \dodoi{10.1051/0004-6361:20065216}

\bibitem[{{Ormel} {et~al.}(2015){Ormel}, {Shi}, \& {Kuiper}}]{Ormel2015}
{Ormel}, C.~W., {Shi}, J.-M., \& {Kuiper}, R. 2015, \mnras, 447, 3512,
  \dodoi{10.1093/mnras/stu2704}

\bibitem[{{Paardekooper} {et~al.}(2023){Paardekooper}, {Dong}, {Duffell},
  {Fung}, {Masset}, {Ogilvie}, \& {Tanaka}}]{Paardekooper2023}
{Paardekooper}, S., {Dong}, R., {Duffell}, P., {et~al.} 2023, in Astronomical
  Society of the Pacific Conference Series, Vol. 534, Protostars and Planets
  VII, ed. S.~{Inutsuka}, Y.~{Aikawa}, T.~{Muto}, K.~{Tomida}, \& M.~{Tamura},
  685, \dodoi{10.48550/arXiv.2203.09595}

\bibitem[{{Papaloizou} \& {Lin}(1989)}]{Papaloizou1989}
{Papaloizou}, J.~C.~B., \& {Lin}, D.~N.~C. 1989, \apj, 344, 645,
  \dodoi{10.1086/167832}

\bibitem[{{Piso} \& {Youdin}(2014)}]{Piso2014}
{Piso}, A.-M.~A., \& {Youdin}, A.~N. 2014, \apj, 786, 21,
  \dodoi{10.1088/0004-637X/786/1/21}

\bibitem[{{Pollack} {et~al.}(1996){Pollack}, {Hubickyj}, {Bodenheimer},
  {Lissauer}, {Podolak}, \& {Greenzweig}}]{Pollack1996}
{Pollack}, J.~B., {Hubickyj}, O., {Bodenheimer}, P., {et~al.} 1996, \icarus,
  124, 62, \dodoi{10.1006/icar.1996.0190}

\bibitem[{{Samsing} {et~al.}(2022){Samsing}, {Bartos}, {D'Orazio}, {Haiman},
  {Kocsis}, {Leigh}, {Liu}, {Pessah}, \& {Tagawa}}]{Samsing2022}
{Samsing}, J., {Bartos}, I., {D'Orazio}, D.~J., {et~al.} 2022, \nat, 603, 237,
  \dodoi{10.1038/s41586-021-04333-1}

\bibitem[{{Shakura} \& {Sunyaev}(1973)}]{SS1973}
{Shakura}, N.~I., \& {Sunyaev}, R.~A. 1973, \aap, 24, 337

\bibitem[{{Sirko} \& {Goodman}(2003)}]{Sirko2003}
{Sirko}, E., \& {Goodman}, J. 2003, \mnras, 341, 501,
  \dodoi{10.1046/j.1365-8711.2003.06431.x}

\bibitem[{{Stevenson}(1982)}]{Stevenson1982}
{Stevenson}, D.~J. 1982, \planss, 30, 755, \dodoi{10.1016/0032-0633(82)90108-8}

\bibitem[{{Stone} {et~al.}(2020){Stone}, {Tomida}, {White}, \&
  {Felker}}]{Stone2020}
{Stone}, J.~M., {Tomida}, K., {White}, C.~J., \& {Felker}, K.~G. 2020, \apjs,
  249, 4, \dodoi{10.3847/1538-4365/ab929b}

\bibitem[{{Tagawa} \& {Haiman}(2023)}]{Tagawa2023}
{Tagawa}, H., \& {Haiman}, Z. 2023, \mnras, 526, 69,
  \dodoi{10.1093/mnras/stad2616}

\bibitem[{{Tagawa} {et~al.}(2020){Tagawa}, {Haiman}, \& {Kocsis}}]{Tagawa2020a}
{Tagawa}, H., {Haiman}, Z., \& {Kocsis}, B. 2020, \apj, 898, 25,
  \dodoi{10.3847/1538-4357/ab9b8c}

\bibitem[{{Thompson} {et~al.}(2005){Thompson}, {Quataert}, \&
  {Murray}}]{Thompson2005}
{Thompson}, T.~A., {Quataert}, E., \& {Murray}, N. 2005, \apj, 630, 167,
  \dodoi{10.1086/431923}

\bibitem[{{Wang} {et~al.}(2022){Wang}, {Jiang}, {Shen}, {Ho}, {Vestergaard},
  {Ba{\~n}ados}, {Willott}, {Wu}, {Zou}, {Yang}, {Wang}, {Fan}, \&
  {Wu}}]{Wang+2022}
{Wang}, S., {Jiang}, L., {Shen}, Y., {et~al.} 2022, \apj, 925, 121,
  \dodoi{10.3847/1538-4357/ac3a69}

\bibitem[{{Wang} {et~al.}(2024){Wang}, {Lin}, {Zhang}, \& {Zhu}}]{WangYH2024}
{Wang}, Y., {Lin}, D. N.~C., {Zhang}, B., \& {Zhu}, Z. 2024, \apjl, 962, L7,
  \dodoi{10.3847/2041-8213/ad20e5}

\bibitem[{{Xu} {et~al.}(2018){Xu}, {Bian}, {Shen}, {Zuo}, {Fan}, \&
  {Zhu}}]{Xu2018}
{Xu}, F., {Bian}, F., {Shen}, Y., {et~al.} 2018, \mnras, 480, 345,
  \dodoi{10.1093/mnras/sty1763}

\bibitem[{{Xu} {et~al.}(2025){Xu}, {Jiang}, {Kunz}, \& {Stone}}]{Xu2025}
{Xu}, W., {Jiang}, Y.-F., {Kunz}, M.~W., \& {Stone}, J.~M. 2025, \apj, 986, 91,
  \dodoi{10.3847/1538-4357/add14a}

\bibitem[{{Xu} \& {Kunz}(2021)}]{Xu2021}
{Xu}, W., \& {Kunz}, M.~W. 2021, \mnras, 502, 4911,
  \dodoi{10.1093/mnras/stab314}

\end{thebibliography}
\bibliographystyle{aasjournal}



\end{CJK*}
\end{document}